\documentclass[aps,pra,twocolumn,showpacs,floatfix,nofootinbib,groupedaddress]{revtex4}

\usepackage{graphicx}
\usepackage{hyperref}
\usepackage{amsmath}
\usepackage{amssymb}

\def\be{\begin{equation}}
\def\ee{\end{equation}}
\def\bea{\begin{eqnarray}}
\def\eea{\end{eqnarray}}
\def\>{\rangle}
\def\<{\langle}

\begin{document}

\title{Measurement Induced Localization of Relative Degrees of Freedom}

\author{Hugo Cable}
\email[]{hcable@www.phys.lsu.edu}
\affiliation{QOLS, Blackett Laboratory, Imperial College London, Prince Consort Road, London. SW7 2BW UK}
\author{Peter L. Knight}
\affiliation{QOLS, Blackett Laboratory, Imperial College London, Prince Consort Road, London. SW7 2BW UK}
\author{Terry Rudolph}
\affiliation{QOLS, Blackett Laboratory, Imperial College London, Prince Consort Road, London. SW7 2BW UK\\
Institute for Mathematical Sciences, Imperial College London, 53 Exhibition Road, London. SW7 2BW UK}

\date{November 11th, 2005}


\begin{abstract}
We present a comprehensive study, using both analytical and numerical methods, of measurement-induced localization
of relational degrees of freedom.
Looking first at the interference of two optical modes, we find that the localization of the relative
phase can be as good for mixed states --- in particular for two initially Poissonian or thermal states --- as for the well-known case
of two Fock states. In a realistic setup the localization for mixed states is robust and experimentally accessible,
and we discuss applications to superselection rules.
For an ideal setup and initial Fock states we show how a relational Schr\"{o}dinger cat state emerges,
and investigate circumstances under which such a state is destroyed.
In our second example we consider the localization of relative atomic
phase between two Bose Einstein condensates, looking particularly at the build up of spatial interference patterns, an area which has attracted much attention since the work of Javanainen and Yoo. We show
that the relative phase localizes much faster than was intimated in previous studies
focusing on the emerging interference pattern itself.
Finally, we explore the localization of relative spatial parameters discussed in recent work
by Rau, Dunningham and Burnett. We retain their models of indistinguishable scattering but make different assumptions. In particular we
consider the case of a real distant observer monitoring light scattering off two particles, who
records events only from a narrow field of view. The localization is only partial regardless of the number of observations.
This paper contributes to the wider debate on relationism in quantum mechanics,
which treats fundamental concepts --- reference frames and conservation laws ---
from a fully quantum and operational perspective.
\end{abstract}

\maketitle

\section{Introduction}
\label{sec:Introduction}

It is a generally accepted principle of modern physics that absolute physical quantities
have no intrinsic usefulness or physical relevance. While the issues had been debated for
centuries, they found their modern expression with Mach, whose influence on Einstein
during his formulation of General Relativity is part of physics mythology. Understanding
to what extent various theories are (or can be made) completely relational (Machian) is,
however, a somewhat slippery business. For example, the Machian features of General
Relativity were elucidated most clearly by Barbour and co-workers years after Einstein's
original publication \cite{barbour}.

Part of the problem when trying to examine issues of relationalism in physics, is that we
generically are forced to describe our physical surroundings in terms of some specific
\emph{reference frame}. A reference frame is simply a mechanism for breaking some
symmetry, and if we are careful then we need to describe the reference frame itself in
terms of the specific physical objects of which it is comprised. For internal
self-consistency, this procedure should be undertaken within the confines of the physical
theory under examination. Once we have done so, it is perilously easy to describe physics
once again in ``absolute'' terms - properties look absolute with respect to the one fixed
reference frame.  To avoid this pitfall one common procedure is to examine the
\emph{translation} of the physical description from one observer's reference frame to
another, and such translation yields insight into the relational features of the physics
under consideration.

In applying this sort of thinking to quantum mechanics several problems present themselves
fairly quickly. The first is that of setting up a reference frame described in
purely quantum mechanical terms. The extent to which this is a problem depends upon the
extent to which one is prepared to accept classical objects and fields within the theory.
Opinions vary. At one extreme classical clocks, spatial reference frames and the like
are simply presumed to exist; the quantum mechanical systems under investigation are
taken to couple to the classical reference frames in such a way that, for example, ``position of the
object'' is by fiat well defined after an appropriate measurement. This is
the common perspective taken when teaching wave-mechanics for instance. At the other
extreme, popular in certain approaches to quantum gravity,
every reference frame object (clock, pointer, etc) is assigned a quantum
mechanical state. Issues surrounding the macroscopic limit (or otherwise) of these
objects must then be tackled.

The second main set of problems encountered quantum mechanically are related to dynamics.
Issues of the specific dynamical couplings between the objects under investigation and
the objects comprising the frame become important, most notably the effects of
``backreaction''. The specific dynamics involved are also of importance in trying to examine
how one might translate between physical descriptions of the same system by two different
observers. The inevitable disturbances that arise in quantum mechanical procedures one
observer may implement in order to fix a system with respect to their frame, generically
force a dynamical examination of translation into a different observer's frame (as
opposed to the kinematical translation possible in classical theories).

In this paper we examine some simple dynamical mechanisms wherein some relationally
defined degree of freedom of two systems becomes well localized (in some sense
``classical'') with respect to some specific observer. In particular we consider two
systems initially uncorrelated with respect to some relative degree of freedom, and
examine processes whereby an observer may induce a correlation. We focus on measurement
based schemes, i.e. situations wherein the observer seeks to establish correlation by
appropriate measurement. In contrast to most situations when studying quantum measurements,
wherein the information obtained about the pre-measurement state is the priority, here we
are interested in controlling the properties of the induced post-measurement state. This
sheds light on certain process whereby an observer may use one system as a reference for
another. We examine the speed at which these references are created, and the stability of the
relationship once established. We look at cases wherein the initial states of the systems
under consideration are mixed, in addition to the more commonly considered pure state
case.

A final set of issues of interest in examining relationalism
in quantum mechanics involves conservation laws, superselection
rules and symmetry breaking. In algebraic quantum field theory, the existence of absolute
conservation laws and associated superselection rules (rules forbidding the creation of
superpositions of states with different values of the conserved quantity) is taken to be
true axiomatically \cite{haag}. A less absolutist, and more operational, approach was
initiated by Aharonov and Susskind \cite{aharonov}. They suggested that forbidden
superpositions can in fact be observed provided that the apparatus used by an observer
are prepared in certain special states. The states suggested by
Aharanov and Susskind were not particularly realistic. We present as an alternative
certain mixed states with well localized relative phase,
which are much more experimentally feasible, and can
reproduce the desired effects with no loss due to the lack of purity.

We begin in Sec.~\ref{sec:RLopticalphase} by considering the
localization of relative optical phase, given two cavities of
photons initially in Fock (number) states. Aspects of this problem
were analyzed numerically by M{\o}lmer \cite{molmerone,molmertwo}
and analytically by Sanders et al \cite{sanders}.
In an ideal setup a ``relational Schr\"{o}dinger cat state'' emerges,
and we discuss how slight imperfections lead to destruction of the cat.
We progress to the
mixed state case investigating the localization between two
Poissonian or two thermal states, introducing a two mode visibility
for rigorous comparison. In particular, we show that the
localization of the relative phase of the mixed states is just as good
as for the pure states in these cases.

In Sec.~\ref{sec:RLatomicphase} we look at the interference of two Bose condensates, a process
in which relative localization of atomic phase plays an important role. A numerical
analysis was performed by Javanainen and Yoo \cite{javanainenone,javanainentwo},
and some analytic analysis was given by Castin and Dalibard
\cite{castin}. In our analysis we borrow from our study of localizing optical phase
and see in particular that the localization of atomic phase takes place on the same rapid
time scale, and much faster than is apparent in the simulated spatial interference
of Javanainen and Yoo.

Finally, in Sec.~\ref{sec:RLposition}, we turn to the localization of
relative position, a problem studied recently by Rau et al
\cite{rau}. We extend their results to scenarios involving initially mixed
states and less specialized scattering processes which generate only
partial localization.

\section{Localization of relative optical phase}
\label{sec:RLopticalphase}
\subsection{Pure initial states}
\label{sec:Fockstates}

\begin{figure}
\includegraphics[height=6cm]{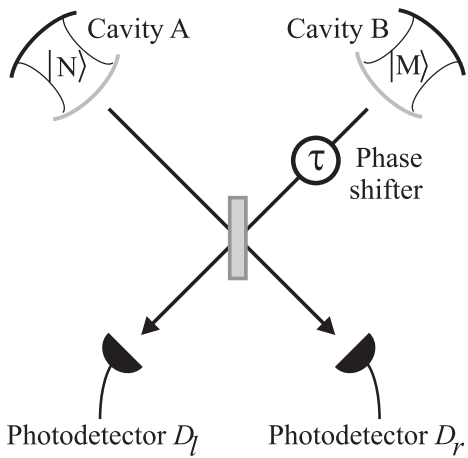}
\caption{\label{fig:twocavities} Photon number states leak out of their cavities and are combined
on a 50:50 beamsplitter. The two output ports are monitored by photodetectors. The variable phase
shift $\tau$ is initially fixed at 0.}
\end{figure}

We begin our study of relative localization in quantum mechanics by examining in detail
the dynamical localization of the relative phase of two, initially independent, single
modes of light. A simple operational procedure for both causing and probing such
localization is depicted in Fig.~\ref{fig:twocavities}. Two cavities
initially containing $N$ and $M$ photons respectively (and thus described by pure initial
states $|N\>|M\>$) both leak out one end mirror (via linear mode coupling). Their outputs
are combined on a 50:50 beamsplitter, after which they are detected.

Despite the cavities being in Fock states with no well-defined relative phase, it is well
known that an interference pattern is observed at the two detectors. The interference
pattern can be observed in time if the two cavities are populated by photons of slightly
differing frequencies or, as in standard interferometry, by varying a phase shifter
placed in one of the beamsplitter ports. The reason for this contradiction with the naive
dictum ``number and phase are conjugate quantities'' may be understood as follows:

Consider the case after a single photon has been detected at one of the detectors. Then
the new state of the two cavities is
\[
\sqrt{\tfrac{N}{N\!+\!M}}|N\!-\!1\>|M\> \pm \sqrt{\tfrac{M}{N\!+\!M}}|N\>|M\!-\!1\>\,,
\]
i.e. it is entangled.
It is simple to show that the second photon is much more likely to be detected at
the same detector. The exact ratio of the probabilities of being counted at the same detector and
at the other is $N^2+M^2-N-M+4NM$ to $N^2+M^2-N-M$.
When $N=M$ this ratio is strictly greater than $3$, and tends sharply to infinity as $N$ and $M$ approach $1$.  This is in
agreement with the phenomenon, demonstrated by the well-known Hong, Ou and Mandel dip experiment \cite{HongOuMandel},
whereby two uncorrelated and identical photons, simultaneously incident on the input ports of a $50:50$ beamsplitter, must
both be registered at the same output port.  Further detections lead to a more and more entangled
state. It is not so surprising then that detections on an entangled state lead to some form of
interference pattern. In essence, after a small number of detections the relative number
of photons in each cavity is no longer well defined, and so a well defined relative phase
can emerge. Note that this is only possible if the beamsplitter, detectors and cavities all
have well defined relative positions.

One method for confirming this intuition is to use  a quantum jumps approach
(for a review of quantum jump methods see \cite{PlenioKnight} and references therein), and
numerically simulate such a system through a number of detection procedures
as was performed in \cite{molmerone,molmertwo}.
However such simulations yield little in the way of physical insight. As
such, we follow instead a procedure introduced in \cite{sanders}. We begin by expanding
the initial state $|\psi_I\>=|N\>|M\>$ of the cavities in terms of coherent states
$|\alpha\>,|\beta\>$:
\be
|\psi_I\>=\mathcal{N}\int_0^{2\pi} \!\!\!\int_0^{2\pi}
\!\!\!d\theta d\phi e^{-i(N\theta+M\phi)} |\alpha\>|\beta\>
\ee
with $\alpha=\sqrt{N}e^{i\theta}$, $\beta=\sqrt{M}e^{i\phi}$, and the normalization
$\mathcal{N}=1/\sqrt{\Pi_N(N) \Pi_M(M)}{4\pi^2}$  where $\Pi_n(\mu)=\mu^n e^{-\mu}/n!$ is the
Poissonian distribution. For the moment we will ignore normalization.

Consider now the case that a single photon is detected at either the left detector $D_L$,
or the right one $D_R$. Since we are interested only in the change of state in the cavity
modes, we treat the exterior modes as ancillas, and find the Kraus operators $K_L,K_R$
describing the effect of the detection (for an explanation of quantum operations see, for example,
\cite{NielsenChuang}). It is reasonably simply to verify that they are
proportional to $a\pm b$, where $a,b$ are annihilation operators for the modes in cavity
$A,B$ respectively. The constant of proportionality depends on the transmittivity of the
end mirrors.

In the event that some number $l$ of photons are detected in $D_L$ while $r$ photons are
detected in $D_R$, the state of the two cavities evolves as follows:
\begin{eqnarray*}
|\psi_I\>&\rightarrow& K_L^l K_R^r |\psi_I\>\\
 &\propto& K_L^l K_R^r \int \!\!\!\int\!\!\!d\theta d\phi e^{-i(N\theta+M\phi)} |\alpha\>|\beta\> \\
 &\propto& \int \!\!\!\int\!\!\!d\theta d\phi e^{-i(N\theta+M\phi)} (\alpha-\beta)^l (\alpha+\beta)^r  |\alpha\>|\beta\>
\end{eqnarray*}
In order to understand the localization in relative phase which occurs between the two
cavities, we need to consider the co-efficient
\be
C_{l,r}\equiv (\alpha-\beta)^l (\alpha+\beta)^r.
\ee
For our purposes it is sufficient to focus on the case that the cavities begin in the same
photon number state; it should be noted however that the physics of the highly asymmetric
case is somewhat different.

\begin{figure}
\includegraphics[height=6cm]{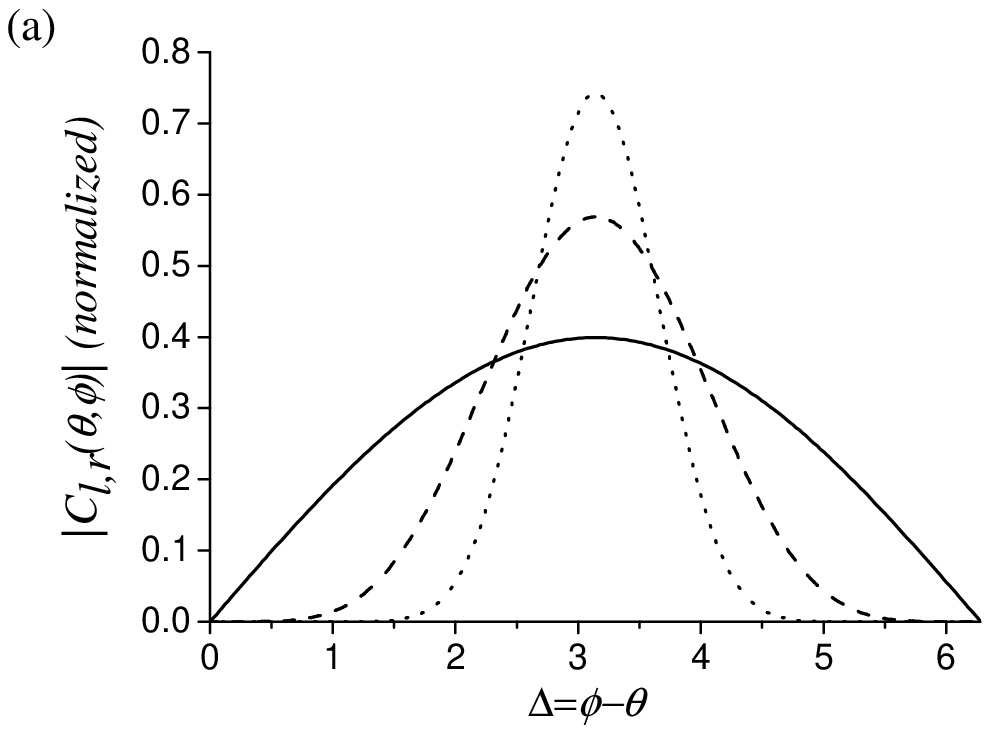}
\includegraphics[height=6cm]{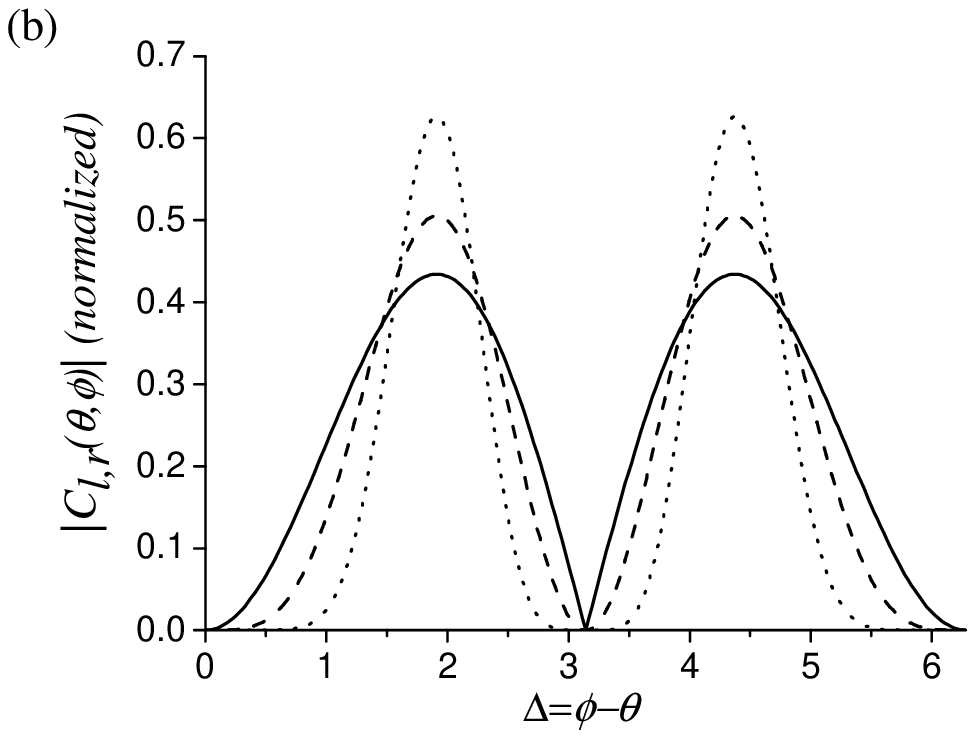}
\caption{\label{fig:RLoptical} The evolution of $C_{l,r}(\theta,\phi)$.
In (a) localization about $\Delta_0=\pi$ after 1, 5 and 15 counts when photons are recorded in the left photodetector only.
(b) localization about $\Delta_0 = \pm 2\arccos \left( 1/\sqrt{3} \right) \sim 1.9$ after 3, 6 and 15 counts when twice as many photons are recorded in the left detector as the right one. The symmetry properties of the Kraus operators $K_L$ and $K_R$ cause
$C_{l,r}$ to have multiple peaks.
}
\end{figure}

Consider first the case that $|\alpha|=|\beta|$, that is, $N=M$. In this case,
\begin{eqnarray}
\label{eqn:Clrcavitiesequal}
C_{l,r}(\theta,\phi)
\!\!\! &=& \!\!\! N^{(l+r)/2} (e^{i\theta}-e^{i\phi})^l (e^{i\theta}+e^{i\phi})^r \\
\!\!\! &=& \!\!\! (4N)^{(l+r)/2} (-i)^l e^{i(l+r)(\theta+\phi)/2}  \sin^l {\! \tfrac{\Delta}{2}} \cos^r \! \tfrac{\Delta}{2} \nonumber
\end{eqnarray}
where $\Delta\equiv (\phi-\theta)$. For the moment we can ignore factors that do not
depend on $\theta,\phi$, since they will be taken care of by normalization.

We are particularly interested in the behavior of $C_{l,r}(\theta,\phi)$ as the total number
$l+r$ of detections gets larger.
To examine this limit, we make use of asymptotic expansions \cite{rowe}
for $C_{l,r}(\theta,\phi)$ as follows.
When photons are detected at both detectors,
\begin{eqnarray}
\label{eqn:bothportsasympt}
| \sin^{l} \tfrac{\Delta}{2} \cos^{r} \tfrac{\Delta}{2} |
\! &\approx& \!
\sqrt{\frac{l^{l}r^{r}}{ l\!+\!r \,^{l+r}}}\,{\exp} \left[ - \! \tfrac{l\!+\!r}{4} \! \left( \Delta \!-\! \Delta_0 \right) ^{2} \right]
\end{eqnarray}
where $\Delta_0 \equiv 2 \arccos\sqrt{r/(r+l)}$ when $\Delta$ takes values between $0$ and $\pi$,
and $\Delta_0 \equiv 2\pi \! - \!  2 \arccos\sqrt{r/(r+l)}$ between $\pi$ and $2\pi$.
$\Delta_0$ denotes the values of the relative phase around which the localization occurs.
When all the photons are detected at one detector the appropriate expressions are
\begin{eqnarray}
\label{eqn:oneportasympt}
\left\vert \cos ^{r}\tfrac{\Delta }{2}\right\vert
\! &\approx& \!
{\exp} \left[ -\tfrac{r}{8}\Delta ^{2} \right] \;\;\;{\rm for}\; \Delta \in [-\pi,\pi], \nonumber \\
\left\vert \sin ^{l}\tfrac{\Delta }{2}\right\vert
\! &\approx& \!
{\exp} \left[ -\tfrac{l}{8}\! \left( \Delta \!-\! \pi \right) ^{2} \right] \;\;\;{\rm for}\; \Delta \in [0,2\pi].
\end{eqnarray}

We see that as $l+r$ gets larger, the state of the two cavities evolves into a
superposition (over global phase) of coherent states with a increasingly sharply defined
relative phase. A plot showing the evolution of $C_{l,r}(\theta,\phi)$ is shown in
Fig.~\ref{fig:RLoptical}. This localization in relative phase is
responsible for the interference phenomena seen at the two detectors, as was examined
numerically by M{\o}lmer \cite{molmerone,molmertwo}.

This is our first concrete example of dynamical relative localization, and so we explore
carefully the key features. Firstly, the value $\Delta_0$ at which the relative phase
localization occurs depends on the (ratio of) the specific number of photons $l,r$
detected at each detector. This is of course probabilistic and we denote by $P_{l,r}$ the
probability of detecting $l$ and $r$ photons in the left and right detectors respectively.
A complete expression for $P_{l,r}$ is obtained by a simple heuristic treatment of the dynamics,
as we now show.

We suppose as in \cite{sanders} that population leaks out of each cavity according to a linear coupling with parameter
$\epsilon$ and that $\epsilon$ is small. After the action of the beam splitter and the
photon detections, $l$ at the left detector and $r$ at the right, the full expression for the
cavity modes is,
\[
 \mathcal{N} \!\! \int \!\! d\theta d\phi e^{-iN(\theta +\phi )}C_{l,r}(\epsilon ,\theta ,\phi )\left\vert \sqrt{1-\epsilon }\alpha \right\rangle \left\vert \sqrt{1-\epsilon }\beta \right\rangle
\]
where the normalisation factor $\mathcal{N} = \frac{1}{\Pi_{N}(N)4\pi ^{2}}$ and
\[
 C_{l,r}(\epsilon ,\theta ,\phi )
 =
 \left\<r \, \Big{\vert} \sqrt{\epsilon }\frac{\alpha +\beta }{\sqrt{2}}\right\rangle
 \left\<l \, \Big{\vert} \sqrt{\epsilon }\frac{-\alpha +\beta }{\sqrt{2}}\right\rangle
\]
extracting the $l$ and $r$ photon components of the coherent states.
The probability $P_{l,r}$ is given by,
\[
\mathcal{N}^{2} \!\! \int \!\! d\theta d\theta ^{\prime }d\phi d\phi ^{\prime }e^{-\frac{i}{2}\left( 2N-r-l\right) (\theta +\phi )}e^{\frac{i}{2}\left( 2N-r-l\right) (\theta ^{\prime }+\phi ^{\prime })}
\]
\[
\times \,\, C_{l,r}(\epsilon ,\theta ,\phi )C_{l,r}(\epsilon ,\theta ^{\prime },\phi ^{\prime })^{\ast }
\]
\[
\times \,\, \left\langle \sqrt{1-\epsilon }\alpha ^{\prime }\right\vert \left\vert \sqrt{1-\epsilon }\alpha \right\rangle \left\langle \sqrt{1-\epsilon }\beta ^{\prime }\right\vert \left\vert \sqrt{1-\epsilon }\beta \right\rangle
\]
By treating the coherent states as quasi-orthogonal (for the basic properties of coherent states see, for example, \cite{GerryKnight}),
\begin{eqnarray*}
\left \langle \alpha' \vert \alpha \right \rangle &=& \exp\left(-|\alpha-\alpha'|^2\right) \sim \delta (\phi-\phi') \\
\left \langle \beta' \vert \beta \right \rangle &=& \exp\left(-|\beta-\beta'|^2\right) \sim \delta (\theta-\theta')
\end{eqnarray*}
and using the relation for the gamma function $\Gamma(.)$,
\[
\int_{0}^{2\pi } \!\!\! \int_{0}^{2\pi }\tfrac{d\theta }{2\pi }\tfrac{d\phi }{2\pi }\cos ^{2r} \tfrac{\Delta }{2} \sin ^{2l} \tfrac{\Delta}{2} = \frac{\Gamma (r+0.5)\Gamma (l+0.5)}{\pi \Gamma (r+l+1)}
\]
we obtain the following approximation for $P_{l,r}$,
\be
\label{eqn:FockProbApprox}
P_{l,r} \! \approx \!\!
\left[ \frac{(2\epsilon N)^{r+l}}{\left( r+l\right) !}e^{-2\epsilon N} \! \right] \!\!
\frac{\left( r+l\right) !}{r!l!}\frac{\Gamma (r+0.5)\Gamma (l+0.5)}{\pi \Gamma (r+l+1)}.
\ee
The approximations would na\"{\i}vely be expected to hold good when several, but not too many, photons have been recorded so that $C_{l,r}$ is narrow while the amplitudes $\sqrt{1-\epsilon}\alpha, \sqrt{1-\epsilon}\beta$ are still large.
In fact detailed inspection of the probabilities $P_{l,r}$ computed numerically reveal that
the fractional error of the approximation (\ref{eqn:FockProbApprox}) compared to the exact values is
roughly $0.6\epsilon$, growing linearly with the leakage parameter.
In terms of its general features, $P_{l,r}$ is seen to be a product
of a global Poissonian distribution in the total number of detected photons $l+r$
and a second function depending on the precise ratio of counts at $D_l$ and $D_r$.

\begin{figure}
\includegraphics[height=8cm]{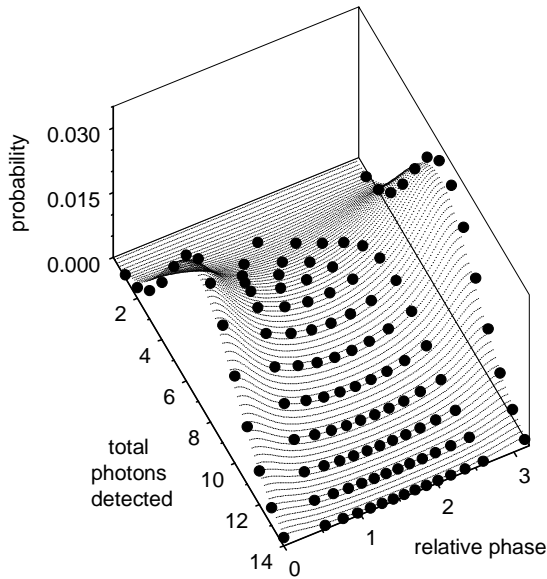}
\caption{\label{fig:FockPlr}
A plot of the exact values of the probabilities $P_{l,r}$ for all the possible measurement outcomes to the procedure
a finite time after the start, against the absolute value of the relative phase which is evolved.
The initial state is $\left\vert 20 \right\rangle \left\vert 20 \right\rangle$ and the leakage parameter $\epsilon$,
corresponding roughly to the time, has a value of $0.2$.  Each spot corresponds to
a different measurement outcome with $l$ and $r$ counts at detectors $D_l$ and $D_r$ respectively.
The value $\Delta_0$ of the relative phase which evolves in each case is given by $2 \arccos\sqrt{r/(r+l)}$.}
\end{figure}

A plot of the exact values for the probabilities $P_{l,r}$ of different measurement records is plotted
in Fig.~\ref{fig:FockPlr} for typical parameter values, $\epsilon=0.2$ and initial state
$\left\vert 20 \right\rangle \left\vert 20 \right\rangle$, where each spot marks a possible measurement outcome.
Plotting $P_{l,r}$ reveals the likely degree of localization of the relative phase $\Delta$
a finite time after the start of the procedure, and the values of $\Delta_0$ which are picked out.
$\epsilon$ corresponds to a time parameter, an approximation which holds good provided $\epsilon$ is not too large.
Looking at the precise distribution in Fig.~\ref{fig:FockPlr}
we see that given $\epsilon=0.2$ it is most likely that 7 photons (approximately $2 \epsilon N$) have been counted,
corresponding to the ridge.  The most probable events involve all the photons being counted at one detector
or the other, picking out a relative phase of $0$ or $\pi$.
However the density of points is greatest about $\Delta_0=\frac{\pi}{2}$ where there are equal counts
at both detectors. Overall no particular value of the relative phase is preferred in this example.

Once a given measurement outcome has occurred with $l$ and $r$ counts in the left and right detectors
respectively the resultant state of the two cavities has two symmetries as can be seen from the explicit
form of $C_{l,r}$, Eq.~(\ref{eqn:Clrcavitiesequal}) and in Fig.~\ref{fig:RLoptical}. A $2\pi$ translational symmetry
identifies physically identical phases. In addition there is symmetry in $C_{l,r}$ about $\Delta=0$. This exists
because the procedure as described so far localizes the absolute value of the relative phase.
When photons are detected at both ports $C_{l,r}$ is peaked at two different values $\pm \Delta_0$.
Looking at the asymptotic form of $C_{l,r}$ as $l$ and $r$ tend to large values we see that
the state that emerges, $|\psi_\infty\>$, takes the following form:
\begin{eqnarray}
\label{eqn:psiinfinitydouble}
|\psi_\infty\>
\!\! &\propto& \!\!
\int \! d\theta e^{-i2|\gamma|^2\theta}
|\gamma\> \nonumber \\
\!\! && \!\!
\otimes \left [
e^{\!-i|\gamma|^2 \! {\Delta_0}} |\gamma e^{i{\Delta_0}}\>
\!+\!
 e^{i|\gamma|^2 \! {\Delta_0}}|\gamma e^{\!-i{\Delta_0}}\>
\right ]
\end{eqnarray}
where $\vert\gamma\>=\vert|\gamma|e^{i\theta}\>$ and $|\gamma|=\sqrt{N-\left(\frac{l+r}{2}\right)}$.
The relative component of the two mode state, contained in the square brackets,
is a superposition of two coherent states with the same amplitude
but different phases $\pm\Delta_0$ - ordinarily called a Schr\"{o}dinger cat state. $|\psi_\infty\>$ has
in addition a sum over all values of the global phase $\theta$. A state of the form Eq. (\ref{eqn:psiinfinitydouble})
could be a termed a \textit{relational Schr\"{o}dinger cat state}.

Creating the superposition Eq.~(\ref{eqn:psiinfinitydouble}) would however be experimentally challenging
as it requires perfect phase stability. In practise we find that the Schr\"{o}dinger cat is sensitive
to any asymmetry, or instability in the system.
The effect of a randomly varying phase is to cause localisation about
\textit{one} particular value of the relative phase. This phenomenon is evident in the numerical studies of Molmer \cite{molmerone,molmertwo}.
These incorporate a slight frequency difference between the two cavity modes causing the free evolution to have
an additional detuning term $\exp i(\omega_b - \omega_a) b^{\dagger}b t$.  Combined with the random intervals
between detections, this means that the process can be described by Kraus operators $a \pm e^{i\tau}b$
where the phase $\tau$ takes random values for each photodetection. The relative phase then takes a unique value
varying randomly for each run. A dynamically equivalent process occurs when atoms
from two overlapping Bose Einstein condensates drop onto an array of detectors and are detected at random positions;
a detailed discussion of this point follows in Sec.~\ref{sec:RLatomicphase}.

In the case of an idealized setup, in which the phase shifts throughout the apparatus remain fixed,
one component of the relational Schr\"{o}dinger cat state can be removed manually.
We suppose that after $l$ and $r$ photons have been detected at $D_l$ and $D_r$ in the usual way
the phase shifter is adjusted by $\pm \Delta_0$ and that the experiment is continued until a
small number of additional photons have been detected. The phase shift translates the interference pattern
in such a way that with high probability the additional counts will occur at one detector.
These additional measurements eliminate the unwanted component of the cat state and confirm
a well defined relative phase.

The next important feature we turn to concerns the robustness of the localization. In the limit of a
large number of detections, the state of the two cavities becomes equivalent to
\be
\label{eqn:psiinfinity}
|\psi_\infty\>=\int \!\! d\theta e^{-2i|\gamma|^2\theta} |\gamma\>|\gamma e^{i{\Delta_0}}\> \\
\ee
with $|\gamma\>=||\gamma|e^{i\theta}\>$ some coherent state. The coherent states, being
minimum uncertainty gaussian states, are the most classical of any quantum states. Thus
we expect states of the form $|\gamma\>|\gamma e^{i{\Delta_0}}\>$ to be robust.
However, $|\psi_{\infty}\>$ is a superposition over such states, and this could
potentially affect the robustness. That this is not the case, can be understood by noting
that the superposition in Eq. (\ref{eqn:psiinfinity}) is summed over the global phase
$\theta$\footnote{By ``global phase'' we are not
referring to the always insignificant total phase of a wavefunction, but rather the phase
generated by translations in photon number: $e^{ia^\dagger a}$. This is still a relative
phase between different states in the Fock state expansion of a coherent state.} of the
coherent states. Under evolutions obeying an additive conservation of energy rule
(photon-number superselection), which is essentially the extremely good rotating-wave approximation of
quantum optics, this global phase becomes operationally insignificant. This is discussed
in a little more detail in Sec.~\ref{sec:superselection} below.

Finally we point out that a state of the form Eq.~(\ref{eqn:psiinfinity}) is, for any processes
involving relative phases between the cavities, operationally equivalent to a tensor
product of pure coherent states for each cavity $|\gamma\>|\gamma e^{i\Delta_0}\>$.
However, because of the phase factor $e^{-2i|\gamma|^2\theta}$, the state is, in
fact, highly entangled. If we expand it in the (orthogonal) Fock bases, as opposed to the
non-orthogonal coherent states, we find a state of the form
\begin{eqnarray}
\left\vert \psi _{\infty }\right\rangle \!\!\!\!
&=& \!\!\!\!
\int \tfrac{d\theta }{2\pi }e^{-2i\left\vert \gamma \right\vert ^{2}\theta }\left\vert \gamma \right\rangle \left\vert \gamma e^{i{\Delta_0}}\right\rangle \nonumber \\
&=& \!\!\!\!
\!\!\! \sum^{\infty}_{n,m=0} \!\!\! \!\! \sqrt{\Pi_{n} \! \left( |\gamma |^{2}\!\right) \! \Pi_{m} \! \left( |\gamma |^{2}\!\right) }
\!\!\int\!\! \tfrac{d\theta}{2\pi} e^{i\left(\! n+m-2\left\vert \gamma \right\vert ^{2}\!\right) \theta }e^{im{\Delta_0}} \! \left\vert n,\!m \!\right\rangle \nonumber \\
&=& \!\!\!\!
\! \sum^{2|\gamma|^2}_{m=0} \!\! \sqrt{\Pi_{2\left\vert \gamma \right\vert^{2}\!-\!m} \! \left( |\gamma |^{2}\! \right) \! \Pi_{m} \! \left( |\gamma |^{2} \! \right) } \: e^{im{\Delta_0}} \! \left\vert 2 \! \left\vert \gamma \right\vert ^{2}\!\!-\!m,\!m\!\right\rangle
\end{eqnarray}
where $\left\vert n,\!m \! \right\rangle$ denotes a product of photon number states, $\Pi_.(.)$
denotes a Poissonian factor and $2|\gamma|^2$ is a whole number of photons.

\subsection{Mixed (poissonian) initial states}
\label{sec:opticalPoissonian}

The example of the previous section, while usefully illustrating many features of
relative localization, is not experimentally accessible due to the assumption
that we have access to large
photon number, initially pure, Fock states populating the cavities. In particular, if we
are looking for a mechanism by which relative localization occurs naturally in our
interactions with surrounding objects, the previous example is somewhat implausible as it
stands, in as much as it would suggest that macroscopic levels of entanglement are
necessary to localize relative degrees of freedom.

With this in mind, we turn to a more realistic scenario. While it is implausible that the
cavities are populated by large Fock states, it is \textit{not} implausible that they are
populated by a large number of photons, and that all we know is the mean number $\bar{N}$
of photons. In such a situation we would assign the quantum state of the cavity as a
Poissonian distribution over photon number (if we were following a maximum entropy
principle). Alternatively, we may be populating the cavities by (independent) light from
lasers, in which standard laser theory leads to the photon number distribution being
Poissonian \cite{scully}.

As such, we reconsider the above localization procedure, assuming now that the initial
state of the cavities is
\begin{eqnarray}
\label{eqn:Poissonianstart}
\rho_I&=&\sum_n \Pi_n(\bar{N})|n\>\<n| \otimes \sum_m \Pi_m(\bar{N})|m\>\<m| \nonumber \\
&=& \frac{1}{4\pi^2}\int\!\!\!\int \!\!d\theta d\phi \;|\alpha\>\<\alpha|\otimes
|\beta\>\<\beta|,
\end{eqnarray}
where $\alpha=\sqrt{\bar{N}} \exp i\theta$ and $\beta=\sqrt{\bar{N}} \exp i\phi$.
As in the previous section, we consider the evolution of $\rho_I$ given that $l,r$
photons are detected at the left and right detectors respectively:
\begin{eqnarray*}
\rho_I&\Rightarrow& K_L^l K_R^r \rho_I K_L^{\dagger l} K_R^{\dagger r} \\
 &\propto& K_L^l K_R^r \int\!\!\!\int \!\!d\theta d\phi \;|\alpha\>\<\alpha|\otimes
|\beta\>\<\beta| K_L^{\dagger l} K_R^{\dagger r} \\
 &\propto& \int \!\!\!\int\!\!\!d\theta d\phi \; |\alpha+\beta|^{2r} |\alpha-\beta|^{2l} |\alpha\>\<\alpha|\otimes
|\beta\>\<\beta| \\
&\propto& \int \!\!\!\int\!\!\!d\theta d\phi \; |C_{l,r}(\theta,\phi)|^2
|\alpha\>\<\alpha|\otimes |\beta\>\<\beta|
\end{eqnarray*}
with $C_{l,r}(\theta,\phi)$ as in Eq.~(\ref{eqn:Clrcavitiesequal}). Clearly the discussion about the
localising nature of $C_{l,r}(\theta,\phi)$ applies
equally well in this case. The expression Eq.~(\ref{eqn:FockProbApprox}) approximating the probabilities
for different measurement records when the initial state is a product of Fock states is exact
for a product of Poissonian states. In the limit of a large number of detections,
\be
\label{eqn:rhoinfinity} \rho_\infty =  \int\!\!\!d\theta \;
|\alpha\>\<\alpha|\otimes |\alpha e^{i\Delta_0}\>\<\alpha e^{i\Delta_0}|
\ee
We see that quite remarkably \emph{the relative phase localization of the mixed states is
just as sharp and just as rapid as that of the pure states}!

\begin{figure}
\includegraphics[height=5cm]{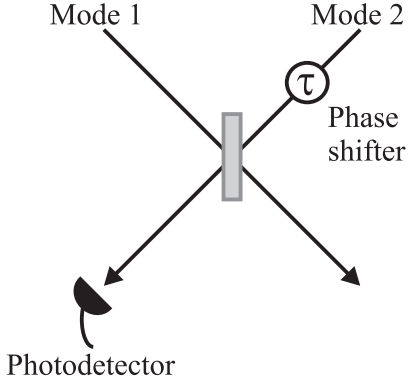}
\caption{\label{fig:visibilitydef}
$I(\tau)$ is the intensity at the left output port after the second
mode undergoes a phase shift of $\tau$ and is combined with the first at a 50:50 beam splitter.
This intensity is evaluated for all possible settings of the phase shifter.
Extremizing over $\tau$, the visibility for the two mode state is defined as
$V=(I_{\rm max}-I_{\rm min}) / (I_{\rm max}+I_{\rm min})$.}
\end{figure}

To rigorously quantify the degree of localization of relative phase we define a visibility
for the prepared two mode state. This definition is illustrated by Fig.~\ref{fig:visibilitydef}.
It is supposed that the second mode undergoes a phase shift $\tau$ before being completely
combined with the first at a 50:50 beam splitter. The expected photon number at the left port
is then denoted $I(\tau)$.  This intensity is evaluated for all possible phase shifts $\tau$,
allowing a visibility for the two mode optical state to be defined in terms of the difference
between the maximum and minimum values as follows,
\be
V=( I_{\rm max} - I_{\rm min} ) \, / \, ( I_{\rm max} + I_{\rm min} ).
\ee
By definition, the visibility takes values between $0$ and $1$.

For a product of photon number states $|N\rangle|M\rangle$ the action of a phase shifter on the second mode
merely introduces an irrelevant factor of $e^{iM}$, and hence the intensity $I(\tau)$ is constant for different phase shifts $\tau$
and the visibility $V$ is $0$.  In a similar way the visibility is $0$ for any product of mixed states
diagonal in the photon number basis, such as the product of Poissonian states in Eq.~(\ref{eqn:Poissonianstart}).
On the other hand, for a product of coherent states $|\sqrt{\bar{N}}e^{i\theta}\rangle|\sqrt{\bar{N}}e^{i(\theta+\Delta_0)}\rangle$
and, $|\psi_\infty\>$ Eq.~(\ref{eqn:psiinfinity}) and $\rho_\infty$ Eq.~(\ref{eqn:rhoinfinity}) summed over the global phase,
and all three with exactly one value $\Delta_0$ for the localized relative phase, we can easily show that
$I(\tau)$ is proportional to $\cos^2 \left( \frac{\Delta_0+\tau}{2} \right)$.  $I(\tau)$ is then maximized if the phase shifter is set to
$\tau=-\Delta_0$ and $0$ for $\tau=-\Delta_0+\pi$.  Therefore the visibility is $1$ for these three examples for which
the relative phase is perfectly correlated.

After involved calculation, extending methods and results developed earlier in this section,
a simple expression for the intensity and the visibility can be found for the case of
two initial Poissonian states and an idealized experiment for which the phase shifts
throughout the apparatus remain fixed (see Appendix \ref{appendix:PoissonianVisibilityDerivation} for the
derivation):
\be
\label{eqn:Poissonianintensity}
I(\tau )\propto r\cos ^{2}\tfrac{\tau }{2}+l\sin ^{2}\tfrac{\tau }{2}+\tfrac{1}{2}
\ee
\be
\label{eqn:Poissonianvisibility}
V_{l,r} = \frac{|r-l|}{r+l+1}.
\ee
If the detections are all at one detector, the right one say,
Eq.~(\ref{eqn:Poissonianvisibility})
simplifies to $r/\left(r+1\right)$ which tends rapidly $1$, and in fact is $1/2$ even after one detection.
However, it is also seen that the expression diminishes to $0$
for measurement outcomes in which the proportion of counts in the left and right detectors becomes equal.
This does not reflect less localization in those cases but is an artefact of the definition of the visibility.
It is easy to see that if the state of the two cavities is localized at two values of the relative
phase these will both contribute to the intensity at one port in the definition of the visibility; changing the phase shift
$\tau$ will tend to reduce the contribution of one while increasing that of the other so that overall the variation in
the intensity is reduced.  However, in realistic situations
we expect the localisation at multiple values to be killed by slight asymmetries
(as in the pure state case), and the visibility to tend to $1$ in all cases.

\subsection{Mixed (thermal) initial states}
\label{sec:opticalthermal}

In the examples presented above, there was no limit to how sharply the localization could
be achieved. We now examine the case that both cavities are initially populated by
thermal states with equal mean photon numbers $\bar{N}$ along the same lines as above.
As such, the initial state is
\begin{eqnarray*}
\rho_I
\!\!&=&\!\!
\sum_n \frac{\bar{N}^n}{(1+\bar{N})^{n+1}}|n\>\<n|
\otimes
\sum_m \frac{\bar{N}^m}{(1+\bar{N})^{m+1}}|m\>\<m| \\
\!\!&=&\!\!
\frac{1}{4\pi^2\bar{N}^2} \! \int \!\! \int \!\! d\bar{n} d\bar{m} d\theta d\phi
e^{-(|\alpha|^2+|\beta|^2)/\!\bar{N}} |\alpha\>\<\alpha| \! \otimes \! |\beta\>\<\beta|,
\end{eqnarray*}
where $\alpha=\sqrt{\bar{n}}\exp i\theta$ and $\beta=\sqrt{\bar{m}}\exp i\phi$.
Under the measurement of $l,r$ photons at the left and right detectors respectively:
\begin{eqnarray*}
\rho_I \!\!\! &\Rightarrow& \!\!\! K_L^l K_R^r \rho_I K_L^{\dagger l} K_R^{\dagger r} \\
\!\!\! &\propto& \!\!\!\! \int\!\! d^2\!\alpha d^2\!\beta \, e^{\!-\left(\!|\alpha|^2+\!|\beta|^2\!\right)/\!\bar{N}} K_L^l K_R^r |\alpha\>\!\<\alpha|
 \!\otimes\!
|\beta\>\!\<\beta| K_L^{\dagger l} K_R^{\dagger r} \\
\!\!\! &\propto& \!\!\!\! \int \!\! d^2\!\alpha d^2\!\beta \, e^{\!-\left(\!|\alpha|^2+\!|\beta|^2\!\right)/\!\bar{N}} |\alpha\!+\!\beta|^{2r} |\alpha\!-\!\beta|^{2l} |\alpha\>\!\<\alpha| \!\otimes\! |\beta\>\!\<\beta| \\
\!\!\! &\propto& \!\!\!\! \int \!\! d^2\!\alpha d^2\!\beta \, e^{\!-\left(\!|\alpha|^2+\!|\beta|^2\!\right)/\!\bar{N}} |C_{l,r}(\bar{n},\bar{m},\theta,\phi)|^2
|\alpha\>\!\<\alpha| \!\otimes\! |\beta\>\!\<\beta|
\end{eqnarray*}

\begin{figure}
\includegraphics[height=6.5cm]{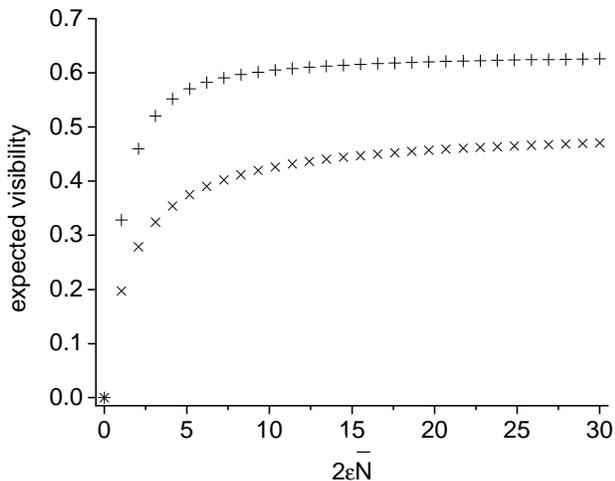}
\caption{\label{fig:mixedstatevisibilities} Expected visibilites for (a) an initial product of two Poissonian states (plusses) and
(b) an initial product of two thermal states (crosses), with average photon number $\bar{N}$ for both cavities.
}
\end{figure}

Unlike previous examples $|C_{l,r}|$ does not provide a simple picture of the localization of the relative
phase due to the additional dependence on the mean photon number variables. However an intensity and a visibility
can be computed as in the Poissonian case above (see Appendix \ref{appendix:ThermalVisibilityDerivation}).
For an arbitrary measurement record the results are,
\be
I(\tau ) \propto l\cos ^{2}\tfrac{\tau }{2}+r\sin ^{2}\tfrac{\tau }{2}+1
\ee
\be
V_{l,r}=\frac{|r-l|}{r+l+2}.
\ee
If all the measurements occur in one detector, the right one say, the visibility is  $r/\left(r+2\right)$
which is $1/3$ after just one detection and which tends to $1$ rapidly
- but slower than in the Poissonian case. In addition an expected visibility
a finite time after the start of the procedure,
$\Sigma_{l,r}P_{l,r}V_{l,r}$,
can be computed using an exact expression for
the probabilities of different measurement records,
\be
P_{l,r}= \frac{(\bar{N}\epsilon )^{r+l}}{\left( 1+\epsilon \bar{N}\right) ^{r+l+2}},
\ee
which notably has the form of the probabilities for two independent sources of thermal light with parameter
$\epsilon \bar{N}$.
The expected visibilities
for the thermal and Poissonian cases are compared in Fig.~\ref{fig:mixedstatevisibilities}.
These averages do not tend to one as the visibility underestimates the degree of localization
when the prepared states are localized at two values of the relative phase,
as discussed previously in Sec.~\ref{sec:opticalPoissonian}.
However the general trend is clear.
The cavity modes initially in thermal states tend, as in the Poissonian case,
to a state which is perfectly correlated in relative phase while remaining
separable.  Although more photons must be detected to achieve the same
degree of localization when the initial states are thermal, the localization
proceeds very rapidly in both cases.

\subsection{The connection to superselection rules}
\label{sec:superselection}

A conservation law makes operational sense (or non-sense) only when related to the
procedures whereby the conserved physical quantities are measured. In particular, the
frame of reference against which the measurements are made plays a crucial role. Certain
frames of reference (e.g. position, atom number) are more in accord with our everyday
experience than others (e.g. ``charge phase'', ``isospin phase''). This has perhaps more
to do with the ground state of the universe (the electromagnetic vacuum in particular)
which acts as a readily accessible reference frame,  than with any fundamental physical
restrictions.

A belief in absolute conservation laws leads to a belief in absolute superselection rules
(SSR). To illustrate how the more relational approach works, let us consider quantum
optics under the ``rotating wave approximation'' (refer for example \cite{GerryKnight}),
equivalent to a strict superselection rule for energy under which the energy is additively
conserved, and in the absence of any absolute phase reference.
Under such an assumption, superpositions of states of
different photon number (such as coherent states of light) and superpositions of
non-degenerate atomic states ($|g\>,|e\>$) are forbidden. Interaction Hamiltonians are
strictly excitation-conserving, for example the familiar Jaynes-Cummings Hamiltonian
$H=|e\>\<g|\otimes a+|g\>\<e|\otimes a^\dagger$ is allowed, where $a,a^\dagger$ are
annihilation, creation operators for an optical mode (for a discussion of Jaynes-Cummings dynamics
refer for example \cite{GerryKnight}).

If we are asked how to operationally create and verify the existence of a superposition
of atomic states of the form $|g\>+|e\>$, then a simple response is to drive the atom
through a cavity containing a large amplitude coherent state of light for an appropriate length of time,
where the interaction Hamiltonian is $H$ above. Measurement of the atom after exiting the
cavity yields it in the ground state half the time. However, this could be due to the
atom being in a mixed state, and thus to verify a coherent superposition has been
obtained, the atom can be sent through a second cavity, also in a large amplitude coherent state,
after which it can be found that the atom is always in the ground state. This
demonstrates it was actually in a coherent superposition $|g\>+|e\>$ between the two
cavities.

What Aharonov and Susskind noted in \cite{aharonov} was that the two cavities did not, in fact, need
to be in coherent states. In fact a state of the form Eq.~(\ref{eqn:psiinfinity}),
taking the total number of photons to be large, is
operationally just as good as initial coherent states for the purposes of demonstrating
a coherent superposition of atomic states as just described\footnote{AS actually
considered creation of a superposition of a proton (equivalent to $|g\>$) and a neutron
(equivalent to $|e\>$) using coherent states - or otherwise - of negatively charged
mesons (equivalent to photons).}. As noted previously, this state is one of fixed total
energy, and thus there is no violation of the conservation law globally.

An objection to this argument having much physical relevance can be made along the lines
discussed above, namely that states of the form Eq.~(\ref{eqn:psiinfinity}) are not easy to come
by in nature, and in fact are highly entangled. We see, however from the results of
Sec.'s~\ref{sec:opticalPoissonian} and \ref{sec:opticalthermal}, that
a mixed state of the form (\ref{eqn:rhoinfinity}) would do
just as well for the operational demonstration of coherent superposition envisaged by AS.
Such mixed states are much more easily preparable, and would seem to conform more closely
with the type of reference frame states that observers typically prepare.

It is observations like this that lend hope to the idea that such dynamical localization
of relational variables may, in fact, be of significance in obtaining a deeper
understanding of quantum mechanics. The effect is not some fragile phenomenon relying on
pure states. In this regard it is also important to note that the localizing mixed states,
in Sec.'s \ref{sec:opticalPoissonian} and \ref{sec:opticalthermal}, are
manifestly separable - they contain no entanglement. In fact, $\rho_{\infty}$
for both Poissonian and thermal initial states
has the interesting feature of being formally separable, but \emph{not} locally preparable under
a superselection rule (or equivalently lack of a suitable reference frame), a feature
first noted in \cite{rudolphsanders}.

\section{Bose-Einstein condensates and relative localization of atomic phase}
\label{sec:RLatomicphase}

A common, and useful, description of BEC's makes use of a coherent state macroscopic
wavefunction for the condensates. Such a description is generally justified by invoking
standard stories about symmetry breaking (for example, see \cite{PitaevskiiStringari}).

There are several reasons to be suspicious of the standard story. The first is that it
requires a description in terms of a coherent superposition of states with different atom
numbers. If, as appears to be a very good approximation in this universe, atom number is
conserved, then such a description is tricky to justify. A common attempt at such
justification is made along the lines that the BEC is surrounded by a thermal cloud with
which it is exchanging atoms and thus the atom number is undetermined. However such a
process leads only to a mixed state for the BEC, and not the desired pure coherent state.
Secondly, the symmetry breaking is generally invoked by the addition of auxiliary fields
with no clear physical relevance. Finally, the most striking demonstrations of coherence
in BEC's come from interference experiments, as is discussed,
for instance, in \cite{KetterleDurfeeStamperKurn}.
However, such experiments do not require
description via atom-number-violating coherent states, and moreover such a description
places an advocate of such a description in the philosophically precarious position of
writing down quantum mechanical states containing in principle unknowable parameters.

Let us point out that there is a difference between experiments in which a single
condensate is coherently ``cut'' into two parts, and then allowed to re-interfere.
Such interference is trivially obtainable without the use of coherent states (as an optical analogue,
sending a photon Fock state - or
even a thermal state - through a Mach-Zehnder interferometer demonstrates perfect
interference!) Thus we are interested only in the case that the BEC's are
independent.

It is simple to imagine an experiment involving two BEC's that closely follows the
optical scenario described above for photons in cavities. For instance,
two condensates trapped in separate
potential wells may be allowed to slowly tunnel through a barrier. Atoms originating from
different wells can be rendered indistinguishable by mixing at an appropriate beam
splitter. While a standard description of the experiment would utilize interference
between coherent condensate fields $|\psi_1|e^{i\theta}$, $|\psi_2|e^{i\phi}$, the
discussion of the previous section can be carried over to conclude that such a
description is not necessary. In fact, it is less desirable - it violates atom number
conservation, and invokes the use of the (independent and) unknowable phases
$\theta,\phi$, which vary from run to run of the experiment, and should therefore be
correctly incorporated in a quantum mechanical framework by the use of mixed initial
states (leading to a description as in Sec.~\ref{sec:opticalPoissonian} above).

In practise the most striking BEC interference patterns are those which do not involve
leaking of single atoms onto a beam-splitter and detection in one of only two channels,
but rather are those in which spatial diffraction of the initially independent BEC's
occurs, and a spatial interference pattern is measured in the region of overlap, as is
reported in \cite{BECexperiments}. We
therefore extend the discussion of the previous section to this type of experiment.

A quantum jumps approach to showing that coherent state description of interference
between independent BEC's was first used by Javanainen and Yoo
\cite{javanainenone,javanainentwo} but is unnecessary.
They showed that interference patterns emerge even if the atom number
superselection rule is obeyed exactly, and the condensates are initially in
atomic Fock states with the same number of atoms.  This work generated much interest,
although very little was done analytically (a notable exception is \cite{castin}).

We consider the same simplified model of Javanainen and Yoo \cite{javanainenone, javanainentwo},
but instead take the initial states to be mixed.
It is assumed that phase diffusion, the shape of the trapping potential and edge effects can be ignored.
Each condensate corresponds to macroscopic occupation of a single particle mode with momentum $k$
and is described by a second quantised plane-wave field of the form $e^{ikx} b_k$. We assume that
the two condensates are initially in Poissonian states with the same expected atom number $\bar{N}$,
see Eq.~(\ref{eqn:Poissonianstart}), and with opposite momenta $\pm k$. The condensates merge over a linear
array of atom detectors and atoms are detected singly. The combined field operator
$\hat{\psi}=e^{ikx_1}b_k+e^{-ikx_1}b_{-k}$ serves as the measurement operator for a detection at position
$x_1$.

The situation here turns out to be dynamically equivalent to the optical problem discussed in Sec.~\ref{sec:RLopticalphase}
when the cavity modes are initially Poissonianly populated and the second cavity
undergoes random phase shifts between detections, for example
because of a frequency mismatch (see the later part of Sec.~\ref{sec:Fockstates}).
Inspecting the atomic measurement operator $\hat{\psi}$, it is seen
that atomic measurements $\pi/k$ apart are equivalent, and further,
that a detection in $\frac{\pi}{2k} \leq x_1 < \frac{\pi}{k}$ is equivalent to
one in $0 \leq x_1 < \frac{\pi}{2k}$ at $x_1-\frac{\pi}{2k}$ with operator $\hat{\psi}=e^{ikx_1}b_k-e^{-ikx_1}b_{-k}$.
For a mixed state
\[
\rho \propto \int d^2 \alpha d^2 \beta P(\alpha,\beta) | \alpha \>\< \alpha |_k \otimes | \beta \> \< \beta |_{-\!k},
\]
the probability density for measurement at $x_1$ with this periodic identification is proportional to
\begin{eqnarray*}
&& tr \left( e^{ikx_1}b_k\!+\!e^{-ikx_1}b_{-k} \right) \rho \left( e^{-ikx_1}b^{\dagger}_k\!+\!e^{+ikx_1}b^{\dagger}_{-k} \right) \\
&& + \left( e^{ikx_1}b_k\!-\!e^{-ikx_1}b_{-k} \right) \rho \left( e^{-ikx_1}b^{\dagger}_k\!-\!e^{+ikx_1}b^{\dagger}_{-k} \right) \\
&\propto&
\int d^2 \alpha d^2 \beta P(\alpha,\beta)(|\alpha|^2+|\beta|^2).
\end{eqnarray*}
On this reduced range every $x_1$ is equally probable and the problem can be treated by assuming ``left''
and ``right'' Kraus operators, $K_{r,\tau} \propto a+e^{i\tau}b$ and $K_{l,\tau} \propto a-e^{i\tau}b$ with $\tau$ taking a
random value for $\tau$ for each measurement.

To understand the characteristic localization of relative phase for the two condensates it is sufficient to
take half the detections at $\tau=0$ and the rest at $\tau=\pi/2$, the largest difference possible.
There is little advantage working as in \cite{molmerone,molmertwo,castin} with a probability density
for the full measurement record involving information about the precise spatial distribution of the atomic detections.
The commutativity of $K_{l,\tau}, K_{r,\tau}$ allows the process to broken down as convenient. We suppose that there are
$M$ measurements at each of $\tau=0$ and $\tau=\pi/2$. The numbers of ``left'' and ``right'' counts at
are denoted respectively by $l_1,r_1$ for $\tau=0$ and $l_2,r_2$ for $\tau=\pi/2$.

The $2M$ measurements cause the initial state with average atom number $\bar{N}$ for each condensate
\[
\rho_I = \int \tfrac{d\theta d\phi}{4\pi^2} | \alpha \>\< \alpha |_k \otimes | \beta \> \< \beta |_{-\!k},
\]
where $\alpha=\sqrt{\bar{N}}e^{i\theta}$ and $\beta=\sqrt{\bar{N}}e^{i\phi}$
to evolve as
\begin{eqnarray*}
\rho \!\!\! &\rightarrow& \!\!\!
\frac{M!}{r_{2}!l_{2}!}\frac{M!}{r_{1}!l_{1}!}\hat{K}_{r,\frac{\pi}{2}}^{\, r_{2}}\hat{K}_{l,\frac{\pi }{2}}^{\, l_{2}}\hat{K}_{r,0}^{\, r_{1}}\hat{K}_{l,0}^{\, l_{1}}
\,\rho\,
\hat{K}_{l,0}^{\, l_{1}\dagger}\hat{K}_{r,0}^{\, r_{1}\dagger}
\hat{K}_{l,\frac{\pi }{2}}^{\, l_{2}\dagger}\hat{K}_{r,\frac{\pi }{2}}^{\, r_{2}\dagger} \\
\!\!\! &=& \!\!\!
\int \!\! \tfrac{d\theta d\phi}{4\pi^2} \;
|C_{l_{1},r_{1}}^{\tau=0}(\theta ,\phi ) \,
C_{l_{2},r_{2}}^{\tau=\! \frac{\pi}{2}}(\theta ,\phi )|^{2}
\left\vert \alpha \right\rangle \!\! \left\langle \alpha \right\vert_{k}
\!\! \otimes \! \left\vert \beta \right\rangle \!\! \left\langle \beta \right\vert_{-\!k} \!,
\end{eqnarray*}
where,
\[
|C_{l,r}^{\tau }(\theta ,\phi )|^{2}
\!=
\frac{ \left( r\!+ l \right) !}{r!l!} \,
|\cos \left(\! \tfrac{\Delta-\,\tau}{2} \!\right) \! |^{2r} \, |\sin \left(\! \tfrac{\Delta-\,\tau }{2} \!\right) \! |^{2l}
\]
and $\Delta=\phi-\theta$. The peaked function $C_{l,r}$ is familiar from Sec.~\ref{sec:RLopticalphase} and
the phase shift $\tau$ causes a translation.

The general features of the localization are as in the optical analysis, Sec.~\ref{sec:RLopticalphase}.
However the effect of the phase shift is to ensure with high probability that exactly one value $\Delta_0$ for
the relative phase is picked out. This is so even when $M$ is small.
The phenomenon can be understood by careful inspection of the measurement record and with the aid of the asymptotic
expressions for $C_{l,r}(\theta,\phi)$ Eq.~(\ref{eqn:bothportsasympt}) and (\ref{eqn:oneportasympt}).
We consider the cases of $M=3$, $8$ and $15$, looking at the ``likely events'' - defined as those with
probability greater than a equal fraction $1/(M+1)^2$. The probabilities of these events
total $0.9$, $0.8$ and $0.8$ respectively. In every case a unique value of $\Delta_0$ is picked out.

In very many cases - all when $M=3$ and half when $M=15$ - all the detections are of the same sort, all
$K_l$ (or all $K_r$), at $\tau=0$, or $\tau=\frac{\pi}{2}$, or both.
In other words at least one component of
$|C_{l_{1},r_{1}}^{\tau=0}(\theta ,\phi ) \, C_{l_{2},r_{2}}^{\tau=\! \frac{\pi}{2}}(\theta ,\phi )|^{2}$
is of the form $|C_{M,0}|^2$ (or $|C_{0,M}|^2$) which has only one peak and a larger spread then
otherwise. The product $|C_{l_{1},r_{1}}^{\tau=0}(\theta ,\phi ) \, C_{l_{2},r_{2}}^{\tau=\! \frac{\pi}{2}}(\theta ,\phi )|^{2}$
in turn has only one peak and is highly probable. Other probable events are such that
one peak of $|C_{l_{1},r_{1}}^{\tau=0}(\theta ,\phi )|^2$ strongly overlaps with one peak of
$|C_{l_{2},r_{2}}^{\tau=\frac{\pi}{2}}(\theta ,\phi )|^2$.
In short, the phase shift of $\pi/2$ makes it impossible for $|C_{l_{1},r_{1}}^{\tau=0}(\theta ,\phi )|^2$ and
$|C_{l_{2},r_{2}}^{\tau=\frac{\pi}{2}}(\theta ,\phi )|^2$
to strongly reinforce each other at more than one value for the relative atomic phase.

In the limit of a large number of detections,
\be
\rho_\infty =  \int\!\!\!d\theta \;
|\alpha\>\<\alpha|_k \! \otimes |\alpha e^{i\Delta_0}\>\<\alpha e^{i\Delta_0}|_{-\!k}.
\ee
When the relative phase is perfectly defined the atomic detections have a probability density
of $\cos^2 (kx_1 -\Delta_0/2)$ (where $x_1$ is the proper position). However, beginning from
initial states with no relative phase correlation, \emph{the value for the relative phase localises
much faster than it takes for the characteristic spatial interference pattern to become well established}.
The numerical studies of Javanainen and Yoo, for example, simulate interference patterns based on
$1000$ atomic measurements. However,
the dependence in Eq.'s~(\ref{eqn:bothportsasympt}) and (\ref{eqn:oneportasympt}) on
the total number of detections $l+r$ demonstrates that the underlying rate of localisation
is similar to that at either of the two values of $\Delta_0$ which evolve when the phase
$\tau$ is fixed, as in Sec.~\ref{sec:RLopticalphase}.
The scalar function $\vert C_{l_{1},r_{1}}^{\tau=0} C_{l_{2},r_{2}}^{\tau=\! \frac{\pi}{2}} \vert^2$
is well estimated by a Gaussian with width between $\sqrt{\frac{2}{M}}$ and $\frac{2}{\sqrt{M}}$.

\section{Relative localization of position}
\label{sec:RLposition}

The reference frame with which we, as human observers, have the most natural familiarity
is position. A recent article \cite{rau} by Rau, Dunningham and Burnett (RDB) examined localization in
relative position for two massive particles. The initial states chosen for each particle
were momentum eigenstates - the particles are supposed to start off delocalized throughout
a region very much longer than the wavelength of the incident light.
Two simple examples were analyzed numerically using a stochastic approach.
RDB suggested that the localizing process might be extendable to many particles with the
emergent relative positions having the properties of classical vector displacements.

\begin{figure}
\includegraphics[height=5.5cm]{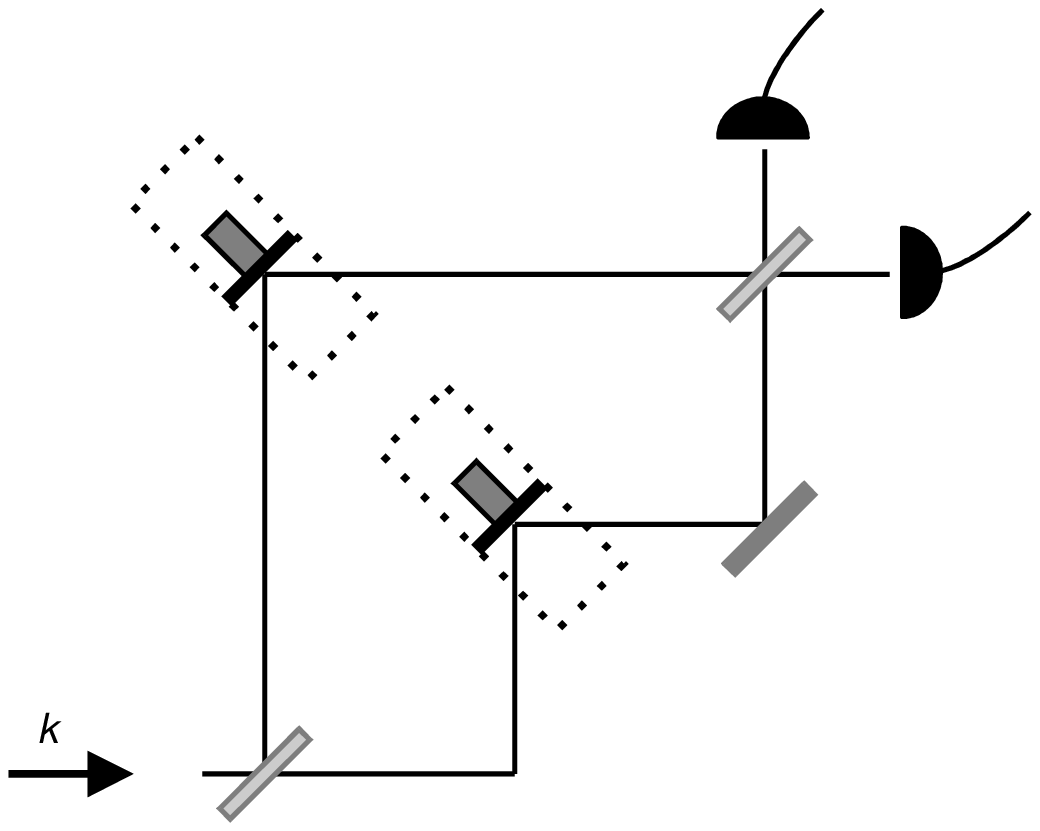}
\caption{\label{fig:rubbercavity}
Photons with momentum $k$ pass through a  ``rubber cavity'' - Mach-Zehnder
interferometer in which two of the mirrors
are mounted on ``quantum springs'' and are initially delocalized along an axis.
Two photodetectors monitor the output channels.}
\includegraphics[height=2.5cm]{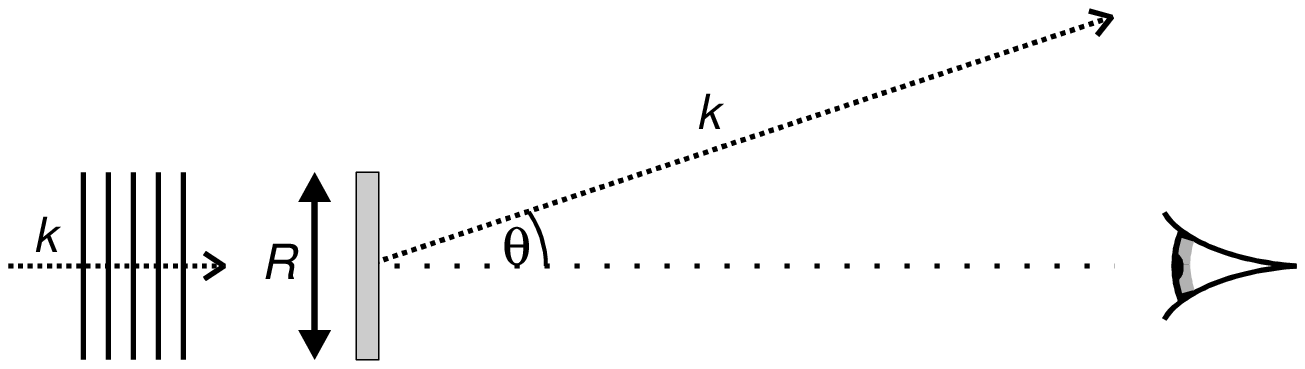}
\caption{\label{fig:freeparticlescattering}
Plane wave photons with momentum $k$ scatter off two free particles, delocalized in a region of length $R$,
and are either deflected at an angle $\theta$ or continue in the forward direction. The observer can
`see' photons which forward scatter or which are deflected only by a small amount.}
\end{figure}

The first example - that of  a ``rubber cavity'' - is illustrated in our Fig.~\ref{fig:rubbercavity} (Fig.~1 of \cite{rau}).
The relative position of two mirrors in a Mach-Zehnder
interferometer is localized by a series of single photons which pass through the device
and are detected by photodetectors monitoring the two output channels.
This example resembles the localization of optical phase
considered in Sec.~\ref{sec:RLopticalphase}. In fact the (numerically produced) Fig.~2 of \cite{rau} is
essentially identical with our (analytic) Fig.~\ref{fig:RLoptical}. The Kraus operators which
summarize the possible outcomes for each photon are proportional to
\be
\label{eqn:rubbercavityoperators}
\exp \left(i\sqrt{2}k \hat{x} \right) \pm \exp \left(i\sqrt{2}k \hat{y}\right)
\ee
where $k$ is
the photon momentum and $\sqrt{2}k$ is the momentum kick imparted
to each mirror, located at $x$ and $y$. Their action in the position basis is analogous to that of the
optical operators $K_l$ and $K_r$ on optical coherent states in Sec.~\ref{sec:RLopticalphase}, although
in the latter case the basis is overcomplete. Differently from the optical case the
pattern of relative spatial localization which emerges has a periodicity of $\pi\sqrt{2}/k$ and
extends throughout the region where the particles were at the start.

The second example considered by RDB is the localization in relative position
(due to the scattering of plane wave photons) off two free particles
boxed in a one dimensional region. Fig.~\ref{fig:freeparticlescattering}
illustrates the situation. The Kraus operators are derived as follows.
It is supposed that each particle
if taken by itself acts as a perfect point scatterer, scattering as S-waves with certainty.
The scattered photons are detected in the far field at some angle $\theta$ of deflection.
This simple scattering cannot yield information about the `position' of the particle, see \cite{Heisenberg}.
Rather each event imparts a variable
momentum kick $k \sin \theta$ with operator $\exp i k \sin \theta \hat{x}$ where $k$ is the momentum of
the incident photon, assumed to approach perpendicularly.
With two particles the Kraus operators are therefore
\be
\label{eqn:freeparticleoperators}
\exp \left( i k \sin \theta \hat{x} \right) + \exp \left(i \phi\right) \exp \left(i k \sin \theta \hat{y}\right),
\ee
supposing that scattering off each particle is indistinguishable. The operators are seen to form a density.
The condition that the two particles should act as a single point scatterer when together sets $\phi=0$.
In addition unitary implies the possibility for forward scattering and the single Kraus operator
is,
\be
\label{eqn:freeparticleoperatortwo}
\sqrt{ \int_0^{2\pi} \tfrac{d\theta}{2\pi} \sin^2 \left( \tfrac{k \sin \theta}{2} (\hat{y}-\hat{x}) \right)}.
\ee
These forward scattering events do contribute to the localizing process and cannot be ignored.

It could be asked why the two particles are not considered to be delocalized in a two dimensional
region.  In fact there is a complication in this case.
Following the same arguments as previously we can easily write down the Kraus operators.
We resolve our vectors in Cartesian components with the ``x-direction'' aligned along the direction
of propagation of the incident plane wave:
\begin{eqnarray*}
\hat{K}_{S} &=& \sqrt{\tfrac{1}{2\pi}}\exp \left( i\Delta k\cdot\hat{m}\right) \cos \left( \tfrac{\Delta k}{2}\cdot\hat{r}\right) \\
\hat{K}_{N} &=& \sqrt{\int_{0}^{2\pi }\tfrac{d\theta}{2\pi} \sin ^{2}\left( \tfrac{\Delta k}{2}\cdot\hat{r}\right) }
\end{eqnarray*}
where the vector position operator for the jth particle is $(x_j,y_j)$,
$\hat{r}=\left( \hat{x}_{2}-\hat{x}_{1},\hat{y}_{2}-\hat{y}_{1}\right) $,
$\hat{m}=\left( \frac{\hat{x}_{1}+\hat{x}_{2}}{2},\frac{\hat{y}_{1}+\hat{y}_{2}}{2}\right) $
and the momentum kick is
$\Delta k=k\left( 1-\cos \theta ,-\sin \theta \right) $.
The problem is this - the operators depend on the vector $\hat{r}$ rather than just $|\hat{r}|$.
They potentially resolve an absolute orientation as well as a relative position with the fixed
direction of the incident photons providing a reference.

The key difference between the rubber-cavity and scattering models, as shown in the numerical
studies of \cite{rau}, is that the changeable momentum kick
of the latter localizes a single value for the relative position rather than a periodic array.
Localization at any value within the initial region is possible.
Notice however that the symmetry about $0$ remains - the scattering process localizes the
absolute value of the relative position leaving two values for the relative displacement, whenever
the initial conditions allow for it.  Comparison should be made with Sec.~\ref{sec:RLatomicphase}
where here translations rather than frequency shifts prevent multiple values for the localization of the relative parameter,
and the multiplicity is eliminated entirely.

In what follows we investigate further the models introduced by RDB. We focus on the case
that the particles share the same attributes and start in the same state.
In particular we use Gaussian states as a basis rather than position eigenstates,
which on their own are dynamically fragile, in as much as they disperse infinitely rapidly under free evolution.
This also facilitates analogy with the localization of relative phase discussed in earlier sections.
The case of localizing relative position turns out to be more technically
complex however, since Gaussian states are not eigenstates of translation operators. In what follows
we must first take a brief diversion to clarify the technicalities of working with Gaussian states.

We employ the notation
$\left\vert \psi _{k,a,d}\right\rangle$ for a Gaussian state with mean momentum $k$, mean position $a$ and spatial spread parameter $d$,
\[
\left\vert \psi _{k,a,d}\right\rangle \propto \sqrt{d} \int_{-\infty }^{\infty }dxe^{ik(x-a)}G_{a,d}(x)\left\vert x\right\rangle
\]
where $G_{a,d}$ denotes a Gaussian probability distribution with mean $a$ and spread $d$,
\[
G_{a,d}(x)=\frac{1}{d \sqrt{2\pi }}e^{-\frac{1}{2}\left( \frac{x-a }{d}\right) ^{2}}.
\]
and $d$ is suppressed when it is constant for consecutive steps.

We consider the effect of a sequence of localizing kraus operators,
$K(\hat{x}_{1},\hat{x}_{2})=K_{i_{N}}(\hat{x}_{1},\hat{x}_{2})\cdots K_{i_{1}}(\hat{x}_{1},\hat{x}_{2})$,
acting on an arbitrary basis state for the two particles. Rather than
explicit localization in the relative mean position parameter $a_2-a_1$,
the basis states evolves to a superposition as follows,
\begin{eqnarray}
\left\vert \psi _{k_{1},a_{1},d}\right\rangle \! \otimes \! \left\vert \psi _{k_{2},a_{2},d}\right\rangle
\!\!\! &\rightarrow& \!\!\!
\left( \tfrac{1}{2\pi }\right) ^{2} \!\! \int \! dx_{1}dx_{2}dp_{1}dp_{2} \nonumber \\
\!\!\! &\times& \!\!\!
K(x_{1},x_{2})e^{ip_{1}(a_{1}-x_{1})}e^{ip_{2}(a_{2}-x_{2})} \nonumber \\
\!\!\! &\times& \!\!\!
\left\vert \psi _{p_{1}+k_{1},a_{1},d}\right\rangle \! \otimes \! \left\vert \psi _{p_{2}+k_{2},a_{2},d}\right\rangle
\end{eqnarray}
where $K(x_{1},x_{2})$ is evaluated at number values and, unless otherwise specified, the integral is
$\int_{-\infty}^{\infty}...\int_{-\infty}^{\infty}\!.$
We can understand this more simply.
Writing $K(x_{1},x_{2})=e^{+i\zeta \frac{(x_{1}+x_{2})}{2}}C_{\Theta }(\frac{x_{2}-x_{1}}{2})$,
a product of translation to the center of mass (where $\frac{\zeta}{2}$ is the cumulative momentum kick)
and the localizing function $C_{\Theta}(\frac{x_{2}-x_{1}}{2})$, we assume
that the Fourier transform $\widetilde{C}_{\Theta }(p)=\int \! dze^{-ipz}C_{\Theta }(z)$
may be defined. Then the final state is proportional to,
\begin{eqnarray}
\label{eqn:relpositionFT}
\!\!\! &&
e^{i\zeta \frac{a_{1}+a_{2}}{2}}
\!\! \int \!\! dp
e^{i\frac{p}{2}(a_{2}-a_{1})}\widetilde{C}_{\Theta }(p) \nonumber \\
\!\!\! &&
\times \left\vert \psi _{\frac{\zeta }{2}-\frac{p}{2}+k_{1},a_{1}}\right\rangle
\otimes \left\vert \psi _{\frac{\zeta }{2}+\frac{p}{2}+k_{2},a_{2}}\right\rangle.
\end{eqnarray}
We see then that the basis states evolve to an increasingly flat superposition over
relative mean momentum with the phase terms recording the location of the relative spatial maxima.
In particular $\widetilde{C}_{\Theta }(p)\simeq e^{-i\frac{p}{2}\Delta _{0}}$
when $K(\hat{x}_1,\hat{x}_2)$ enforces sharp localization of the relative position
at a single value $\Delta_0$.

We can simplify further by
changing basis, regarding the Hilbert space as a tensor product of spaces
for the center of mass and the relative position, rather than spaces for each
particle,
\[
\left\vert x \right\>\! \otimes \!\left\vert y \right\>
\longleftrightarrow
\left\vert \tfrac{x+y}{2} \right\>_{\mathrm{COM}} \!\! \otimes \! \left\vert \tfrac{y-x}{2} \right\>_{\mathrm{Rel}} \!.
\]
The basis state considered above can be rewritten as another product of Gaussian states,
\begin{eqnarray}
&& \left\vert \psi _{k_{1},a_{1},d}\right\rangle
\! \otimes \!
\left\vert \psi _{k_{2},a_{2},d}\right\rangle \nonumber \\
&\propto&
\left\vert \psi _{k_{1}+k_{2},\frac{a_{1}+a_{2}}{2},\frac{d}{\sqrt{2}}}\right\rangle_{\!\!\mathrm{COM}}
\!\!\! \otimes \!
\left\vert \psi _{k_{2}-k_{1},\frac{a_{2}-a_{1}}{2},\frac{d}{\sqrt{2}}}\right\rangle_{\!\!\mathrm{Rel}}\!\!\!\!. \nonumber
\end{eqnarray}
The final state in the new notation after $K(\hat{x}_1,\hat{x}_2)$ has acted enforcing sharp localization at $\Delta_0$
is proportional to
\begin{eqnarray}
\label{eqn:RLcomrellimit}
&& \!\! e^{i\zeta \frac{a_{1}+a_{2}}{2}}\left\vert \psi _{k_{1}+k_{2}+\zeta ,\frac{a_{1}+a_{2}}{2},\frac{d}{\sqrt{2}}}\right\rangle_{\! \mathrm{COM}} \nonumber \\
&\otimes& \!\! \int \!\! dp \, e^{i\frac{p}{2}(a_{2}-a_{1}-\Delta _{0})}\left\vert \psi _{k_{2}-k_{1}+p,\frac{a_{2}-a_{1}}{2},\frac{d}{\sqrt{2}}}\right\rangle _{\! \mathrm{Rel}}\!\!\!\!.
\end{eqnarray}
The center of mass component is merely translated. The localization in the relative component
is best seen by comparison with the following identity, expanding an arbitrary position
eigenstate in terms of Gaussians:
\[
\int dp \, e^{ip(a-X)}\left\vert \psi _{k+p,a,d}\right\rangle
\propto e^{ik(X-a)}G_{a,d}(X)\left\vert X\right\rangle.
\]
If $X$ is far from $a$ the norm vanishes.

We now turn our attention to the localization of relative position
for two particles as might actually occur in nature. Rather than the
the pure momentum states chosen by RDB for initial states,
we consider the localization between two thermal particles.
It is assumed that the two particles have equal mass $m$ and temperature $T$.
We assume first a localizing process that can pick out one particular
value $\Delta_0$ for the relative position.

A thermal state for one particle is given by a mixture of
momentum eigenstates, weighted according to the classical Maxwell-Boltzmann distribution
and can be expressed in terms of Gaussian states in a simple diagonal form,
\begin{eqnarray}
\label{eqn:MaxwellBoltzmann}
\! && \!\! \int \!\! dp \,
\sqrt{\tfrac{1}{2\pi m k_{B}T}} \exp \left( -\tfrac{p^{2}}{2mk_{B}T}\right)
\! \left\vert p\right\rangle \!\!
\left\langle p\right\vert \nonumber \\
\! &=& \!\!
\int \!\! \tfrac{da}{2\pi}
\left\vert \psi_{0,a} \right\rangle \!\!
\left\langle \psi_{0,a} \right\vert.
\end{eqnarray}
The spatial spread parameter of the Gaussian states is given by $d=\sqrt{\tfrac{1}{2mk_{B}T}}$;
when the particle is heavy and hot the Gaussian
states approximate position eigenstates.
We get rid of the infinite limits in Eq.~({\ref{eqn:MaxwellBoltzmann}}) and work with normalizable
states delocalized over a finite region $R$. For both particles together,
\be
\rho_I \propto \int_{R}\int_{R}da_{1}da_{2}\left\vert \psi _{0,a_{1}}\right\rangle \left\langle \psi _{0,a_{1}}\right\vert \otimes \left\vert \psi _{0,a_{2}}\right\rangle \left\langle \psi _{0,a_{2}}\right\vert.
\ee

Under the action of the localizing process $K(\hat{x}_1,\hat{x}_2)$ the initial state $\rho_I$ is transformed
as follows, keeping with the same notation as introduced previously,
\begin{eqnarray}
\rho_I \!\!\! &\rightarrow& \!\! K_{i_N}\cdots K_{i_2} K_{i_1} \rho K_{i_1}^\dagger K_{i_2}^\dagger \cdots K_{i_N}^\dagger \nonumber \\
&\propto& \int_{R}\int_{R}\int \! da_{1}da_{2}d^{2}x_{1}^{(\prime )}d^{2}x_{2}^{(\prime )}d^{2}p_{1}^{(\prime )}d^{2}p_{2}^{(\prime )} \nonumber \\
&\times&
e^{-i\left\{ p_{1}(x_{1}-a_{1})+p_{2}(x_{2}-a_{2})\right\} }
e^{i\left\{ p_{1}^{\prime }(x_{1}^{\prime }-a_{1}^{\prime })+p_{2}^{\prime }(x_{2}^{\prime }-a_{2}^{\prime })\right\} }
\nonumber \\
&\times& K(x_{1},x_{2})K(x_{1}^{\prime },x_{2}^{\prime })^{\ast } \nonumber \\
&\times&
\left\vert \psi _{p_{1},a_{1},d}\right\rangle \left\langle \psi _{p_{1}^{\prime },a_{1},d}\right\vert \otimes \left\vert \psi _{p_{2},a_{2},d}\right\rangle \left\langle \psi _{p_{2}^{\prime },a_{2},d}\right\vert
\end{eqnarray}
where $d^2x_{1}^{(\prime)}$ abbreviates $dx_{1}dx^{\prime}_{1}$ etc.
When the $K(\hat{x}_1,\hat{x}_2)$ operators enforces sharp localization - at $\Delta_0$ say - the final state takes the simple form
\begin{eqnarray}
\label{eqn:sharplylocalizedparticlesA}
\rho _{\infty }
\!\!\! &\propto& \!\!\!
\int_{R} \!\! \int_{R} \! \int \!
da_{1}da_{2}d^2p^{(\prime)}
e^{\frac{i}{2}(p-p^{\prime })(a_{2}-a_{1}-\Delta _{0})} \nonumber \\
\!\!\! &\times& \!\!\!
\left\vert \psi _{\frac{\zeta }{2}-\frac{p}{2},a_{1}}\right\rangle \!\! \left\langle \! \psi _{\frac{\zeta }{2}-\frac{p^{\prime }}{2},a_{1}}\right\vert
\!\! \otimes \!\!
\left\vert \psi _{\frac{\zeta }{2}+\frac{p}{2},a_{2}}\right\rangle \!\! \left\langle \! \psi _{\frac{\zeta }{2}+\frac{p^{\prime }}{2},a_{2}}\right\vert\!\!.
\end{eqnarray}
The center of mass and relative positions remain
unentangled throughout the localizing process, as is clear for example from Eq.~(\ref{eqn:RLcomrellimit}), and
as would certainly be expected. The two particles however
evolve from being separable to being highly entangled. This contrasts to the localization of relative optical
phase for two initially Poissonian, or thermal optical states which do not become entangled despite
the emergence of strong correlation between them, as discussed in Sec.~\ref{sec:RLopticalphase}.

We now look more closely at the localization in relative position induced by the
scattering operators (\ref{eqn:freeparticleoperators})
and (\ref{eqn:freeparticleoperatortwo}) which describe the general case
of light scattering off two free particles.
We suppose in the first instance that the incident light comes as single photons with
the same frequency. However, differently from RDB, we do not assume a detailed
record for every event. Rather we consider
two types of measurement outcome: a ``forward scattering'' where the incident photon continues
without scattering or is scattered into a small angle between $-\epsilon$ and $\epsilon$; and a ``deflection''
where the photon is scattered outside of this range and the light source dims.
We mix over the possible events constituting a measurement outcome.
This is a reasonable model for a real observer monitoring light from a distant source
scattering off two particles, who only has a limited field of view and cannot
measure the angle of scattering.

As previously we work with thermal particles supposed initially
to be delocalized in a region $R$.
We change basis to separate out the center of mass and relative components.
The initial state of the particles is then,
\[
\rho _{I} \propto \int d\tfrac{a_{1}+a_{2}}{2}
\left\vert \psi _{0,\frac{a_{1}+a_{2}}{2},\frac{d}{\sqrt{2}}}\right\rangle \!\!
\left\langle \psi _{0,\frac{a_{1}+a_{2}}{2},\frac{d}{\sqrt{2}}}\right\vert _{\mathrm{COM}}
\!\!\!\! \otimes \rho_{\,\mathrm{Rel}}
\]
where,
\[
\rho_{\,\mathrm{Rel}}
\! \propto \!
\int_{L_{lower}}^{L_{upper}}
\!\! d\tfrac{a_{2}-a_{1}}{2}
\left\vert \psi _{0,\frac{a_{2}-a_{1}}{2},\frac{d}{\sqrt{2}}}\right\rangle \!\!
\left\langle \psi _{0,\frac{a_{2}-a_{1}}{2},\frac{d}{\sqrt{2}}}\right\vert_{\mathrm{Rel}}\!\!\!,
\]
where having changed integration variables from $a_1$ and $a_2$ to $\frac{a_1+a_2}{2}$ and $\frac{a_2-a_1}{2}$, $L_{lower}$ and $L_{upper}$
denote the lower and upper limits of the inner integral which corresponds to the relative component of the
two particle state.  $L_{lower}$ and $L_{upper}$ depend on the outer integration variable $\frac{a_1+a_2}{2}$, and in effect ensure
that the particles remain within the original region $R$.  With $S$
deflection and $F$ forward-scattering events, $\rho_{\,\mathrm{Rel}}$ evolves to $\rho^{\prime}_{\,\mathrm{Rel}}$ as follows,
tracing out the center of mass component at the end,
\footnote{The momentum kick imparted to the center of mass depends on the angle of scattering
but the linearity of the partial trace procedure ensures that this causes no additional complication.}
\begin{eqnarray}
\rho^{\prime}_{rel} \!\!\!\! &=& \!\!\!
\int \!\! \int \!\!
\int_{L_{lower}\left( \frac{a_{1}+a_{2}}{2}\right) }^{L_{upper}\left( \frac{a_{1}+a_{2}}{2}\right) }
d\tfrac{r}{2} \;\; d\tfrac{r^{\prime }}{2} \;\; d\tfrac{a_{2}-a_{1}}{2} \nonumber \\
\!\!\! &\times& \!\!
G_{\frac{a_{2}-a_{1}}{2},\frac{d}{\sqrt{2}}}(\tfrac{r}{2})G_{\frac{a_{2}-a_{1}}{2},\frac{d}{\sqrt{2}}}(\tfrac{r^{\prime }}{2})
\nonumber \\
\!\!\!\! &\times& \!\!\!
\Bigg[ \sqrt{\int_{0}^{2\pi }d\theta \sin ^{2}\left( \tfrac{k\sin \theta r}{2}\right) }\sqrt{\int_{0}^{2\pi }d\theta \sin ^{2}
\left( \tfrac{k\sin \theta r^{\prime }}{2}\right) } \nonumber \\
\!\!\! && \;\;\;
+\int_{-\epsilon }^{\epsilon }d\theta \cos \left( \tfrac{k\sin \theta r}{2}\right) \cos \left( \tfrac{k\sin \theta r^{\prime }}{2}\right) \Bigg] ^{F}
\nonumber \\
\!\!\!\! &\times& \!\!\!
\left[ \int_{\epsilon }^{2\pi -\epsilon }
\!\!\!\!\!\!\!\!\!\!\!\!
d\theta \cos \left( \tfrac{k\sin \theta r}{2}\right) \cos \left( \tfrac{k\sin \theta r^{\prime }}{2}\right) \right] ^{S}
\!\! \left\vert \tfrac{r}{2}\right\rangle \!\! \left\langle \tfrac{r^{\prime }}{2}\right\vert _{\mathrm{Rel}}\!.
\end{eqnarray}

\begin{figure}
\includegraphics[height=11cm]{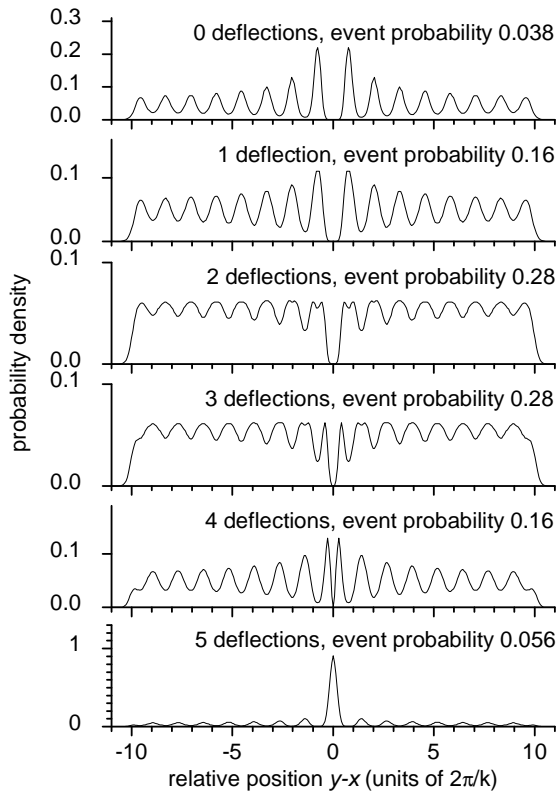}
\caption{\label{fig:monochromaticlightfreeparticles}
Probability densities between $L_{lower}$ and $L_{upper}$ for the relative separation of two free thermal particles after
$5$ photons, each with momentum $k=5$, have scattered off them,
either being deflected into some large angle or continuing in the forward direction.
The spatial spread parameter
$d=\sqrt{\tfrac{1}{2mk_{B}T}}$ is set to $0.2$ (units $2\pi/k$).}
\end{figure}

A typical pattern of localization is shown in Fig.~(\ref{fig:monochromaticlightfreeparticles})
which plots the probability density
$P(y-x)\propto\langle \frac{y-x}{2} \vert \rho^{\prime}_{rel} \vert \frac{y-x}{2} \rangle$ for different
ratios of ``deflection'' and ``forward-scattering'' events, where $x$ and $y$ are the (precise) positions
of each particle; prior to the scattering process $P(y-x)$ is uniformly distributed.
In contrast to the
sharp localization discussed previously,
limited knowledge of the scattering record
means that the localization of relative position is only partial, even after many photons have been scattered.
Rather than peaks in the relative position
we see complex interference patterns.

As expected the degree of localization (and associated probabilities)
for different outcomes are found to be insensitive to the precise value of the small parameter
$\epsilon$ describing the narrow range of angles visible to the observer. Taking $\epsilon=0$,
the patterns are characterized by Bessel functions
of the first kind $P_{\mathrm{Rel}}(y-x)\sim \left[ 1-J_{0}(k(y-x))\right] ^{F}\left[ 1+J_{0}(k(y-x))\right] ^{D}$,
where $D$ denotes the number of deflections and $F$ the number of photons continuing in the forward direction.
The localization is symmetric about the origin.
Sharp localization at one specific value - $y=x$ - is possible with small probability
and occurs when every photon is deflected.

\begin{figure}
\includegraphics[height=11cm]{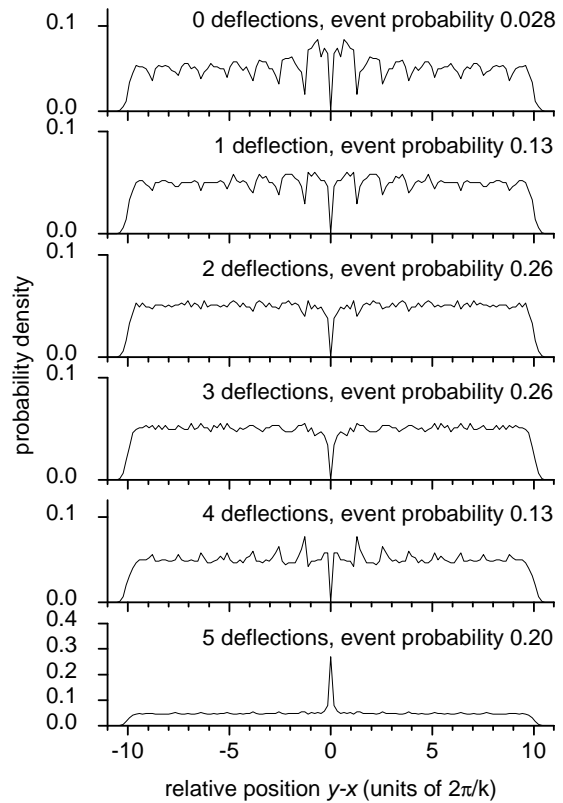}
\caption{\label{fig:thermallightfreeparticles}
Probability densities for the relative separation of two free thermal particles after
$5$ thermal wavepackets have scattered off them,
either being deflected into some large angle or continuing in the forward direction.
The momentum parameter $k=5$ and the spatial spread parameter $d=0.2$ (units $2\pi/k$).
The thermal wavepackets have mean photon number $\bar{n}=5$.}
\end{figure}
In our final example we consider the case that the scattering light is thermal, as well as
the two massive particles. Each incident wavepacket is described by the mixture
$\rho=\Sigma_n \frac{\bar{n}^n}{(1+\bar{n})^{n+1}}\vert n\>\<n \vert$ (where $\vert n\>$ denotes
an $n$ photon Fock state). The scattering operators Eq.~(\ref{eqn:freeparticleoperators}) and
Eq.~(\ref{eqn:freeparticleoperatortwo}) must be modified. We replace the fixed momentum kick
$k\sin\theta$ by the operator $\hat{N} k\sin\theta$ where $\hat{N}$ is the number operator
for the optical mode. Scattering of a single thermal wavepacket leads to a
variable photon count, all detected at a single angle $\theta$ of deflection.
A typical pattern of localization is shown in Fig.~(\ref{fig:thermallightfreeparticles})
which plots the probability density $P(y-x)$ after five thermal wavepackets have scattered;
the results are not sensitive to the precise value of the small parameter
$\epsilon$.

\section{Discussion}

We have examined in depth the measurement induced relative
localization of some interesting quantum mechanical degrees of
freedom. Along the way we have characterized the differences that
occur when mixed states are involved (in particular showing that for
certain mixed states the relative localization can be just as good
as for initially pure states) and provided analytic analyses of
situations which previously have been studied numerically. We have
shown that the relative phase localization in BEC's can be faster
than the interference pattern emergence, a feature not discernible
from the previous numerical studies.

There are certain natural directions in which our results should be
extended. At present we are uncertain whether certain values for the
localizing parameter can be favored (i.e. dominate) after averaging
over all measurement records, particularly if one system is
macroscopic compared to the other. In the context of understanding
the construction of reference frames for quantum systems, this
asymmetric case becomes particularly important. Another question
that becomes important is that of the \emph{transitivity} of the
localisation when applied to more than two systems. For most of the
processes we have studied analytically in this paper it is
reasonably trivial to see that such transitivity does in fact occur.

We have touched on the connection to superselection rules. Two other
areas of fundamental interest that have some connection with the
ideas presented here are symmetry breaking and decoherence. The
former because a symmetry breaking is performed with respect to some
frame of reference which, if treated quantum mechanically, should
amount to a dynamical localization of some relative parameter in a
way that results in final states quantitatively similar to those we
have discussed here. The latter because an approach to decoherence
in which the ``localisation to pointer basis states'' is performed
in a completely relational picture is desirable if one wishes to
extend the ideas of decoherence theory to closed systems (such as
the universe).

%
%
%
%
%
%

\begin{acknowledgments}
This work was supported in part by the UK EPSRC, and by the European Union.
\end{acknowledgments}

\appendix

\section{Derivation of the visibility for Poissonian states}
\label{appendix:PoissonianVisibilityDerivation}

The initial state of the cavity fields, a product of two Poissonian states
both with average photon number $\bar{N}$,
\[
\rho=
\int \frac{d\theta d\phi}{4\pi ^{2}} \left\vert \alpha \right\rangle \left\langle \alpha \right\vert \otimes \left\vert \beta \right\rangle \left\langle \beta \right\vert
\]
where $\alpha=\sqrt{\bar{N}}e^{i\theta}$ and $\beta=\sqrt{\bar{N}}e^{i\phi}$
acquires a factor,
\[
C_{l,r}(\theta ,\phi )=
\left\< r \, {\Big\vert} \frac{\sqrt{\epsilon }\alpha +\sqrt{\epsilon }\beta }{\sqrt{2}}\right\rangle
\left\< l \, {\Big\vert} \frac{-\sqrt{\epsilon }\alpha +\sqrt{\epsilon }\beta }{\sqrt{2}}\right\rangle,
\]
extracting the $l$ and $r$ photon components of the coherent states,
under the canonical localizing processing, in which a fraction $\epsilon \bar{N}$

leaks out of
each cavity and $l$ and $r$ photons are detected at the left and right detectors respectively.
$C_{l,r}$ is peaked at $\pm \Delta_0$ given by $2 \arccos \sqrt{r/r+l}$. The final
state is then,
\[
\rho ^{\prime } \!\!=\!\! \frac{\epsilon ^{r+l}}{4\pi ^{2}r!l!}e^{-2\epsilon \bar{N}}
\!\!\! \int \!\!\!
d\theta d\phi
\!
\left\vert \frac{\alpha +\beta }{\sqrt{2}}\right\vert ^{2r}
\!\!
\left\vert \frac{-\alpha +\beta }{\sqrt{2}}\right\vert ^{2l}
\!\!\!\!
\left\vert \alpha ^{\prime }\right\rangle \!\! \left\langle \alpha ^{\prime }\right\vert \! \otimes \!
\left\vert \beta ^{\prime }\right\rangle \!\! \left\langle \beta ^{\prime }\right\vert
\]
where $\alpha ^{\prime}=\sqrt{1-\epsilon}\alpha$ and $\beta ^{\prime}=\sqrt{1-\epsilon}\beta$.

A probability can be calculated using,
\[
\int d\theta d\phi \cos ^{2r}\frac{\phi-\theta}{2}\sin ^{2l}\frac{\phi-\theta}{2}
= 4 \pi
\frac{\Gamma (r+0.5)\Gamma (l+0.5)}{\Gamma (r+l+1)}
\]
\begin{eqnarray*}
P_{l,r}(\epsilon ,\bar{N}) &=& tr\rho ^{\prime } \\
&=& \frac{(2\epsilon \bar{N})^{r+l}}{r!l!}e^{-2\epsilon \bar{N}}\frac{\Gamma (r+0.5)\Gamma (l+0.5)}{\pi \Gamma (r+l+1)}
\end{eqnarray*}

The visibility of $\rho ^{\prime}$ is computed as follows. The second mode undergoes a
variable phase shift of $\tau$ and both modes are then combined at a $50:50$
beamsplitter. $\rho ^{\prime}$ is goes to $\rho ^{\prime\prime}$ according to,
\[
\left\vert \alpha ^{\prime }\right\rangle \left\langle \alpha ^{\prime }\right\vert \otimes \left\vert \beta ^{\prime }\right\rangle \left\langle \beta ^{\prime }\right\vert \rightarrow
\]
\[
\left\vert \frac{\alpha ^{\prime }\!+\!\beta ^{\prime }e^{i\tau }}{\sqrt{2}}\right\rangle \left\langle \frac{\alpha ^{\prime }\!+\!\beta ^{\prime }e^{i\tau }}{\sqrt{2}}\right\vert \otimes \left\vert \frac{-\alpha ^{\prime }\!+\!\beta ^{\prime }e^{i\tau }}{\sqrt{2}}\right\rangle \left\langle \frac{-\alpha ^{\prime }\!+\!\beta ^{\prime }e^{i\tau }}{\sqrt{2}}\right\vert.
\]
And intensity is then defined as
\begin{eqnarray*}
I(\tau )&=& tr(a^{\dagger }a\rho ^{\prime \prime }) \\
&\propto& \int d\theta d\phi \left\vert \frac{\alpha +\beta }{\sqrt{2}}\right\vert ^{2r}\left\vert \frac{-\alpha +\beta }{\sqrt{2}}\right\vert ^{2l}\left\vert \frac{\alpha ^{\prime }+\beta ^{\prime }e^{i\tau }}{\sqrt{2}}\right\vert ^{2}
\end{eqnarray*}
where the constant of proportionality is of no interest as it divides out when computing the visibility.
Expanding the last term of the integrand,
\[
\left\vert \frac{\alpha ^{\prime }+\beta ^{\prime }e^{i\tau }}{\sqrt{2}}\right\vert ^{2}=2(1-\epsilon )\bar{N}\cos ^{2}(\frac{\Delta +\tau }{2})
\]
where $\Delta=\phi-\theta$.
The expression for $I(\tau)$ may be simplified.
\begin{eqnarray*}
I(\tau ) &\propto& \int d\theta d\phi \cos ^{2r}(\frac{\Delta }{2})\sin ^{2l}(\frac{\Delta }{2}) \\
&& \Big[ \cos ^{2}(\frac{\Delta }{2})\cos ^{2}\frac{\tau }{2}\\
&& -2\cos (\frac{\Delta }{2})\sin (\frac{\Delta }{2})\cos \frac{\tau }{2}\sin \frac{\tau }{2}\\
&& +\sin ^{2}(\frac{\Delta }{2})\sin ^{2}\frac{\tau }{2} \Big]
\end{eqnarray*}
The first and last contributions can be resolved in terms of Gamma functions as for the probability
above, and the second term evaluates to $0$. So,
\[
I(\tau )\propto r\cos ^{2}\frac{\tau }{2}+l\sin ^{2}\frac{\tau }{2}+\frac{1}{2}
\]
and extremizing at $\tau=0$ and $\tau=\pi$,
\begin{eqnarray*}
V&=&\left( I_{\rm max}-I_{\rm min} \right) / \left( I_{\rm max} +I_{\rm min} \right) \\
&=& \frac{|r-l|}{r+l+1}.
\end{eqnarray*}

\section{Derivation of the visibility for thermal states}
\label{appendix:ThermalVisibilityDerivation}

The calculations follow a similar line to the Poissonian case above.
The initial state of the two cavity fields, a product of two thermal states
with the same average photon number $\bar{N}$,
\[
\rho = \left( \frac{1}{\bar{N}\pi }\right) ^{2}
\!\!\! \int \!\! d^{2}\alpha d^{2}\beta \exp \!-\!\left( \frac{|\alpha |^{2}
\!+\!
|\beta |^{2}}{\bar{N}}\right) \left\vert \alpha \right\rangle \!\! \left\langle \alpha \right\vert
\! \otimes \! \left\vert \beta \right\rangle \!\! \left\langle \beta \right\vert
\]
where $\alpha \!=\! \sqrt{\bar{n}}e^{i\theta }$ and $\beta \!=\! \sqrt{\bar{m}}e^{i\phi }$
acquires a factor,
\[
\left\< r \, {\Big\vert} \frac{\sqrt{\epsilon }\alpha +\sqrt{\epsilon }\beta }{\sqrt{2}}\right\rangle
\left\< l \, {\Big\vert} \frac{-\sqrt{\epsilon }\alpha +\sqrt{\epsilon }\beta }{\sqrt{2}}\right\rangle,
\]
extracting the $l$ and $r$ photon components of the coherent states,
under the canonical localizing process.

The final state is,
\begin{eqnarray*}
\rho ^{\prime } \!\! &=& \!\!
\frac{\epsilon ^{r+l}}{\bar{N}^{2}\pi ^{2}r!l!}\int d^{2}\alpha d^{2}\beta
\exp -\left( \frac{|\alpha |^{2}\!+\!|\beta |^{2}}{\bar{N}}\right) \\
&\times& \!\! \left\{\! \exp \!\!-\!\!\epsilon \! \left( \left\vert \frac{\alpha \!+\! \beta }{\sqrt{2}}\right\vert ^{2} \!+\! \left\vert \frac{-\alpha \!+\! \beta }{\sqrt{2}}\right\vert ^{2}\right) \!\right\} \left\vert \frac{\alpha \!+\!\beta }{\sqrt{2}}\right\vert ^{2r}
\!\left\vert \frac{-\alpha \!+\! \beta }{\sqrt{2}}\right\vert ^{2l} \\
&\times& \!\! \left\vert \alpha ^{\prime }\right\rangle \! \left\langle \alpha ^{\prime }\right\vert
\otimes
\left\vert \beta ^{\prime }\right\rangle \! \left\langle \beta ^{\prime }\right\vert
\end{eqnarray*}
where
$\alpha ^{\prime }=\sqrt{1-\epsilon }\alpha $ and $\beta ^{\prime }=\sqrt{1-\epsilon }\beta $.
This expression may be simplified using the parallelogram rule,
\[\left| \frac{\alpha +\beta }{\sqrt{2}} \right|^{2}+\left|\frac{-\alpha +\beta }{\sqrt{2}}\right|^{2}
=|\alpha |^{2}+|\beta |^{2}\]
giving,
\begin{eqnarray*}
\rho ^{\prime } \!\! &=& \!\!
\frac{\epsilon ^{r+l}}{\bar{N}^{2}\pi ^{2}r!l!}\int d^{2}\alpha d^{2}\beta \\
&\times&
\left\{ \exp -\left( \epsilon +\frac{1}{\bar{N}}\right) \left( \left\vert \frac{\alpha +\beta }{\sqrt{2}}\right\vert ^{2}+\left\vert \frac{-\alpha +\beta }{\sqrt{2}}\right\vert ^{2}\right) \right\} \\
&\times&
\left\vert \frac{\alpha +\beta }{\sqrt{2}}\right\vert ^{2r}\left\vert \frac{-\alpha +\beta }{\sqrt{2}}\right\vert ^{2l}\left\vert \alpha ^{\prime }\right\rangle \! \left\langle \alpha ^{\prime} \right\vert
\otimes
\left\vert \beta ^{\prime }\right\rangle
\!\left\langle \beta ^{\prime }\right\vert
\end{eqnarray*}

A probability can be calculated, changing variables of integration such that
$\frac{\alpha+\beta}{\sqrt{2}} \rightarrow \alpha$ and $\frac{-\alpha+\beta}{\sqrt{2}} \rightarrow \beta$
and evaluating with
$\frac{d^{2}\alpha }{\pi } \!=\! d\bar{n} \frac{d\theta }{2\pi }$ and
$\frac{d^{2}\beta }{\pi } \!=\! d\bar{m} \frac{d\phi }{2\pi }$,
\begin{eqnarray*}
P_{l,r}(\epsilon,\bar{N}) \!\! &=& \!\!  tr\rho ^{\prime } \\
&=& \!\! \frac{\epsilon ^{r+l}}{\bar{N}^{2}\pi ^{2}r!l!} \!\! \int \!\! d^{2}\alpha d^{2}\beta \\
&& \times \!
\left\{ \exp \! - \! \left( \epsilon \!+\! \frac{1}{\bar{N}}\right) \left( |\alpha |^{2} \!+\! |\beta |^{2}\right) \right\}
\left\vert \alpha \right\vert ^{2r} \!\! \left\vert \beta \right\vert ^{2l} \\
&=& \!\! \frac{(\epsilon \bar{N})^{r+l}}{\left( 1+\epsilon \bar{N}\right) ^{r+l+2}}
\end{eqnarray*}

The calculation for the intensity $I(\tau)$ for $\rho ^{\prime}$ proceeds as follows,
\begin{eqnarray*}
I(\tau )&=& tr(a^{\dagger }a\rho ^{\prime \prime }) \\
&\propto& \int \frac{d^{2}\alpha }{\pi }\frac{d^{2}\beta }{\pi }\left\{ \exp -\left( \epsilon +\frac{1}{\bar{N}}\right) \left( |\alpha |^{2}+|\beta |^{2}\right) \right\} \\
&& \times \left\vert \alpha \right\vert ^{2r}\left\vert \beta \right\vert ^{2l}
\left\vert \alpha ^{\prime }+\beta ^{\prime }+(-\alpha ^{\prime }+\beta ^{\prime })e^{i\tau }\right\vert ^{2} \\
&\propto& \int d\bar{n}\frac{d\theta }{2\pi }d\bar{m}\frac{d\phi }{2\pi}\left\{ \exp -\left( \epsilon +\frac{1}{\bar{N}}\right) \left( \bar{n}+\bar{m}\right) \right\} \\
&& \times \bar{n}^{r}\bar{m}^{l}\left\vert \sqrt{\bar{n}}e^{i\theta }(1-e^{i\tau })+\sqrt{\bar{m}}e^{i\phi }(1+e^{i\tau })\right\vert ^{2}
\end{eqnarray*}
Now,
\begin{eqnarray*}
&& \left\vert \sqrt{\bar{n}}e^{i\theta }(1-e^{i\tau })+\sqrt{\bar{m}}e^{i\phi }(1+e^{i\tau })\right\vert \\
&=&
\bar{n}|1-e^{i\tau }|^{2}+\bar{m}|1+e^{i\tau }|^{2}+(..)e^{i\theta }e^{-i\phi }+(..)e^{-i\theta }e^{i\phi }
\end{eqnarray*}
and the latter two contributions integrate to $0$. Hence,
\begin{eqnarray*}
I(\tau) \!\! &\propto& \!\! |1-e^{i\tau }|^{2} \!\! \int \!\! d\bar{n}d\bar{m}\bar{n}^{r+1}\bar{m}^{l} \! \exp \!- \left\{ \! \left( \epsilon \!+\! \frac{1}{\bar{N}}\right) \left( \bar{n}\!+\!\bar{m}\right) \! \right\} \\
\!\! &+& \!\! |1+e^{i\tau }|^{2} \!\! \int \!\! d\bar{n}d\bar{m}\bar{n}^{r}\bar{m}^{l+1} \! \exp \!- \left\{ \! \left( \epsilon \!+\!\frac{1}{\bar{N}}\right) \left( \bar{n}\!+\!\bar{m}\right) \! \right\}
\end{eqnarray*}
Evaluating the integrals is as for the probability calculation above.
\begin{eqnarray*}
I(\tau ) &\propto& (|1-e^{i\tau }|^{2}(r+1)+|1+e^{i\tau }|^{2}(l+1) \\
&\propto& l\cos ^{2}\frac{\tau }{2}+r\sin ^{2}\frac{\tau }{2}+1
\end{eqnarray*}
and the visibility is given by
\[ V=\frac{|r-l|}{r+l+2} \]
extremizing at $\tau=0$ and $\tau=\pi$.

\end{document}